\documentclass[sigconf]{acmart}

\usepackage{graphicx} 
\usepackage{nicematrix}
\usepackage{soul}
\usepackage{tikz}
 
\usepackage{amsmath, amssymb, bbm}
 
\usepackage{titlesec}
\usepackage{xspace}
\usepackage{enumitem}
\usepackage{subfigure}
\usepackage{mathtools}
\usepackage{xurl}

\usepackage{soul}
\usepackage{appendix}
\usepackage{float}
\usepackage{sidecap}
\usepackage{pifont}
\usepackage{multicol,wrapfig,multirow}
\usepackage{tablefootnote}

\newcommand{\name}{{\sc Tetris}\xspace}
\newcommand{\teams}{{\sc Teams}\xspace}
\newcommand{\ankur}[1]{}

\newcommand{\rohan}[1]{}

\usepackage[english]{babel}
\usepackage{blindtext}
\renewcommand\footnotetextcopyrightpermission[1]{} 
\setcopyright{none}
\acmDOI{}
\acmISBN{}
\acmPrice{}
\acmConference{}{}{}

\makeatletter
\renewcommand\subsubsection{\@startsection{subsubsection}{1}{\z@}%
  {-0.5\baselineskip \@plus -2\p@ \@minus -.2\p@}%
  {.25\baselineskip}%
  {\ACM@NRadjust\@subsubsecfont}}
\makeatother

\titlespacing\section{3pt}{2pt plus 1pt minus 1pt}{2pt plus 2pt minus 2pt}
\titlespacing\subsection{3pt}{2pt plus 1pt minus 1pt}{2pt plus 2pt minus 2pt}
\titlespacing\subsubsection{3pt}{2pt plus 1pt minus 1pt}{0pt plus 2pt minus 2pt}

\begin{document}
\title{\name: Efficient Intra-Datacenter Calls Packing for Large Conferencing Services}

\author{Rohan Gandhi, Ankur Mallick, Ken Sueda, Rui Liang (Microsoft)}

\begin{abstract}
  Conference services like Zoom, Microsoft Teams, and Google Meet facilitate millions of daily calls, yet ensuring high performance at low costs remains a significant challenge. This paper revisits the problem of packing calls across Media Processor (MP) servers that host the calls within individual datacenters (DCs). We show that the algorithm used in \teams -- a large scale conferencing service as well as other state-of-art algorithms are prone to placing calls resulting in some of the MPs becoming hot (high CPU utilization) that leads to degraded performance and/or elevated hosting costs. The problem arises from disregarding the variability in CPU usage among calls, influenced by differences in participant numbers and media types (audio/video), compounded by bursty call arrivals. To tackle this, we propose \name, a multi-step framework which (a) optimizes initial call assignments by leveraging historical data and (b) periodically migrates calls from hot MPs using linear optimization, aiming to minimize hot MP usage. Evaluation based on a 24-hour trace of over 10 million calls in one DC shows that \name reduces participant numbers on hot MPs by at least $2.5\times$.
\end{abstract}
\maketitle
\sloppy

\section{Introduction}
\label{sec:intro}

Conferencing services such as Zoom\cite{zoom:web}, Google Meet\cite{meet:web}, Microsoft Teams\cite{teams:web}, and DingTalk\cite{dingtalk:web} have become a critical part of the post-COVID19 world. However, the soaring growth in demand\cite{teamsgrowth:web} is also increasing the costs incurred by such service providers. Thus, providing the best user experience at the least cost is an important challenge for such service providers.

Such large-scale conferencing services host multiple millions of calls every day. Each individual call is assigned to a Media Processor (MP) server in cloud data centers (DCs) that receives media streams (such as audio, video, and screen-share) from users, processes, and redistributes the media streams. Getting the call-to-MP assignment right is crucial as it is a key driver of cost and performance\cite{xron:sigcomm23, switchboard:sigcomm23}.

Prior works have focused on assigning calls to MP servers in two phases: (a) select the DC for the individual calls, (b) load balancing (or packing)\footnote{Load balancing and packing are synonymous in this paper.} the calls across MP servers, i.e., assign an MP from the DC selected. Such a division of labor helps scale the MP assignment algorithms to millions of concurrent calls\cite{switchboard:sigcomm23, saving:conext24}. Prior works such as Switchboard\cite{switchboard:sigcomm23} have focused on (a) and have relied on existing state-of-art load balancers for (b).  In this paper, we revisit the problem of assigning calls to MPs in DCs, i.e., phase (b). We found that the algorithm used in Microsoft \teams -- our large scale conferencing service falls short in optimally assigning the MPs to the individual calls. Such an algorithm ends up with large imbalance in the CPU utilization -- some MPs are running with high CPU utilization (hot MPs) while some other MPs are running on low CPU utilization (cold MPs). We find that the problem is not just limited to \teams as other state-of-art load balancing algorithms such as Round-robin, Least-load and others also show the same degree of imbalance. The CPU imbalance translates into either a high degree of overprovisioning (ballooning the costs) or poor user experience (\S\ref{sec:problem}) as more calls get placed on the hot MPs -- neither of which are desired.

There are two unique characteristics of the workload in large scale conferencing services that contribute to the sub-optimal performance of existing load-balancers and make this a novel and challenging scenario:
\begin{enumerate}
    \item \textit{The nature of the calls is such that their CPU utilization changes across different calls.} Calls can have any number of participants. Also, participants use different media streams (audio, video, screenshare), and their quality. All such factors result in CPU usage by the calls vary between calls; two calls can have a vastly different CPU usage. Unfortunately, with existing algorithms such peak demands are not known apriori resulting in higher imbalance (\S\ref{sec:back:workload}). 
    \item \textit{The calls arrive in bursts usually around the 30 and 60 minute marks.} As a result, the \teams controller needs to assign the MPs to a large number of calls simultaneously even when their (peak) demand is unknown.
\end{enumerate}

In this paper, we present \name (Packers and Movers) -- a controller that packs \teams calls across MP servers and improves user experience by reducing calls on hot MPs for a given number of MPs. \name controller runs in each DC and assigns MPs to individual calls in the same DC. 

The design of \name is based on the observation that knowledge of \textit{peak} CPU utilization of a call can improve call assignment to MPs. We can pre-allocate CPU for the peak and help reduce the occurrences of hot MPs. While current algorithms do not attempt to estimate the peak CPU utilization, we show that it is possible to do so with reasonable accuracy and mitigate errors in the estimate through migrations. Specifically, \name is based on three key ideas. Firstly, we recognize that a significant portion (40-60\%) of calls in \teams are recurring, with low variance in participant numbers across instances. Leveraging this, \name utilizes participant data from previous occurrences to estimate the attendance, and consequently peak CPU utilization for the current occurrence. Secondly, for non-recurring calls, a predictive model is employed to estimate maximum participant numbers based on call age and current participants. Although not flawless, both of these predictive models substantially mitigate hot MP instances. Finally, \name reacts to hot MPs by observing that calls last for tens of minutes to hours but stabilize within minutes, prompting migration from hot to cold MPs after an initial period. We model migration as a \textit{bin packing problem} that considers calls and MP servers holistically that can be solved efficiently. We formulate it as a Mixed Integer Program (MIP) and optimize it for scalability and timely execution to mitigate errors in our initial call assignment.

We evaluate \name controller using testbed and simulations using 1-day trace from one of our DCs with O(10 million) calls. Our results show: (a) \name substantially reduces the number of hot MPs, and calls and participants on such hot MPs (\name assigns $2.5\times$ fewer participants on hot MPs compared to state-of-the-art load balancing approaches), (b) the three ideas in \name are effective individually and in combination, (c) benefits of \name continue even on changing the number of MPs in the DC or the fraction of recurring calls, (d) our prediction algorithms have high accuracy, and (e) the MIP provides accurate results in a timely manner at scale.

In summary, the paper makes the following contributions: (a) we show the evidence for high CPU imbalance in \teams -- a large scale conferencing service which translates into poor user experience or high costs, (b) we present \name that uses three complimentary ideas to improve user experience, (c) through at-scale evaluation using a trace from production service, we show \name can substantially reduce the calls and participants on the hot MPs. 
\section{Background}

\subsection{Call assignment in conferencing services}
\label{sec:back:assignment}
%

\begin{figure}[t]
\centering
\includegraphics[width=0.45\textwidth, page=1]{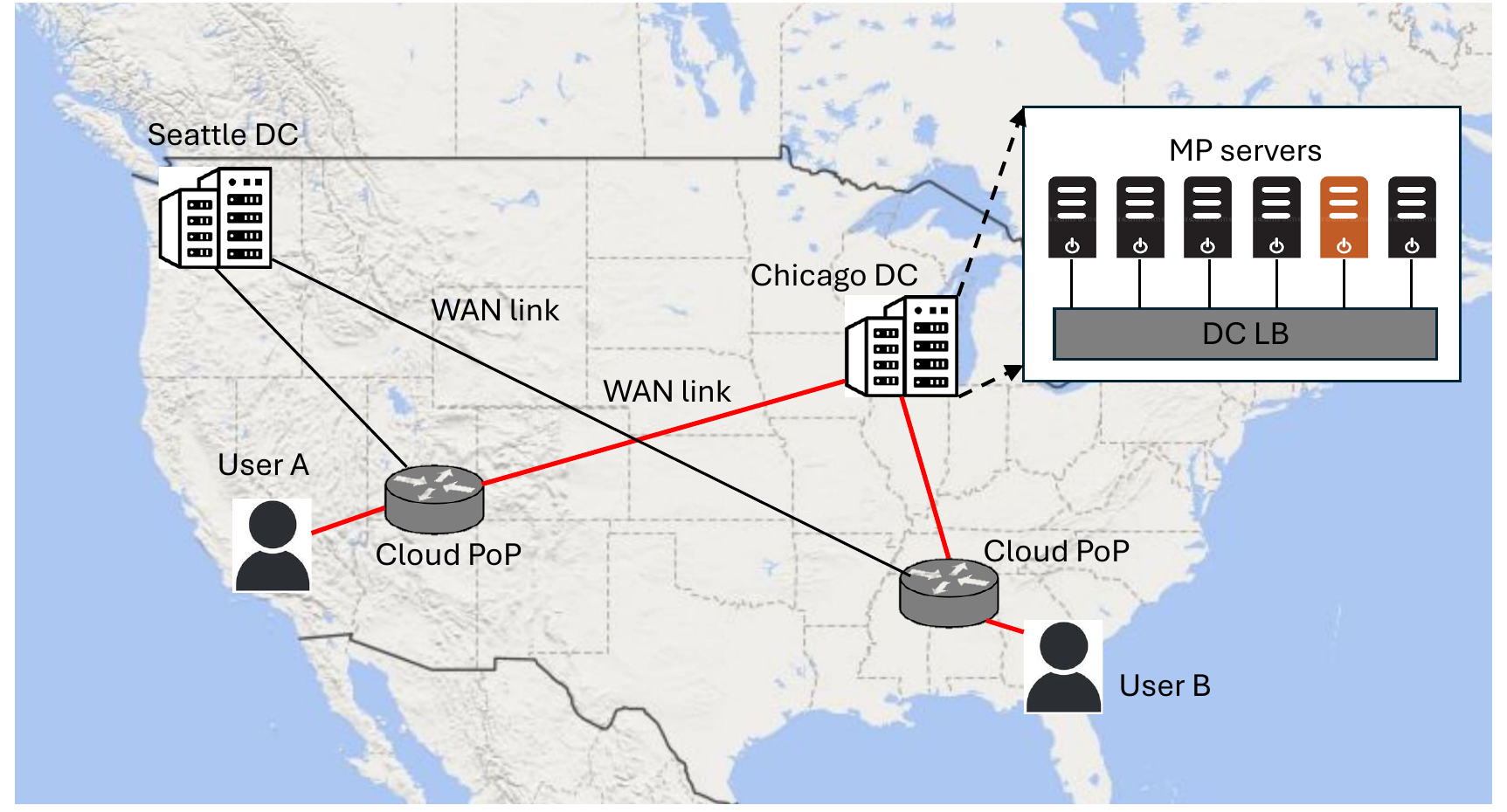}
\caption{Call assignment and routing in \teams. \teams runs in 10+ DCs across the globe. For illustration, we show \teams DCs in Chicago and Seattle. The call between two users (user-A,B) is assigned an orange MP in Chicago DC. Red lines denote the user traffic. Traffic between PoPs and users uses Internet; rest of the traffic is over WAN.}
\vspace{-0.1in}
\label{fig:back:route}
\end{figure}

A large scale conferencing service such as Zoom, Microsoft Teams hosts a large number of simultaneous calls. Each of these calls includes a combination of audio, video and screen share streams. Each of the calls is assigned to a \textit{Media Processor (MP)} server/cluster that receives, processes and re-transmits the streams from/to users\cite{switchboard:sigcomm23, xron:sigcomm23}. Assigning calls to the right MP server is a major challenge for conferencing services. This problem has two aspects\cite{switchboard:sigcomm23, saving:conext24}: 

\textbf{(P1) Selecting the datacenter (DC) for a call.} a large scale conferencing service like \teams is hosted in 10+ DCs across the world for performance, scale and availability reasons. Each DC runs 1000s of MPs. A DC is decided using different policies such as round-robin, locality, or Switchboard\cite{switchboard:sigcomm23}. Fig.\ref{fig:back:route} shows the routing and MP assignment in \teams. The cloud provider based Point-of-Presence (PoPs) receive the traffic. The DC is selected based on the location of the first user of the call. For example, if the first user of the call is from USA, the call is assigned to one of the DCs in USA irrespective of the locations of the subsequent users on the same call. All the subsequent users on that call are directed to the same DC selected based on the location of first user. 

\textbf{(P2) Selecting MP server for a call in a DC.} each DC runs multiple MP servers. Once the DC is determined in P1, the services need to determine the \textit{exact MP server} to host the call in that DC. In other words, we need to load balance calls across MPs. \teams uses algorithm akin to classic load balancer algorithms that aim to spread load across servers (MPs) without causing hotspots. Our algorithm is closest to LLR explained in \S\ref{sec:problem:sim}. As shown in Fig.\ref{fig:back:route}, within the DC, the DC load balancer (DC LB) selects the MPs for the calls. 

\subsection{Functioning of MP}
\label{sec:back:mp}
\textit{Calls with a small or moderate number of participants on the call are assigned to a single MP server}. Such calls account for majority of the calls. Additionally, \textit{these calls also consume significant MP usage.} Large calls, which are handled differently, are outside the scope of this paper. For small to moderate sized calls, the MP receives the media streams (such as audio, video and screen share) from individual participants and distributes them among all participants of that call. The CPU utilization on the MP is mainly driven by the amount of network packets handled (sum of all network packets received and sent out). Note that, the total packets handled is mostly dominated by network packets sent out, which in turn is driven by the number of participants on the call. For example, if there is a call with 5 participants and each participant sends a video of 1 Mbps, then the MP sends out video of 4 Mbps to each participant which adds up to 20 Mbps (in contrast of 5 Mbps of received traffic). 

\textbf{Focus of the paper.} In this paper, we focus on assignment of calls across MPs within the same DC (P2 above). The cost of the MP servers is significant in \teams and making an optimal use of the MP servers is paramount. In addition, keeping CPU utilization low helps reduce queueing lengths and packet drops on the CPU, which in turn reduces video stalls, audio fluency, and user experience\cite{mos:hotnets23}. 

Also, \teams uses Wide-Area-Network (WAN) links to transport traffic between PoPs and DCs (Fig.\ref{fig:back:route}). Akin to past works\cite{saving:conext24}, we have not found WAN to affect the user experience as the WANs are provisioned with enough capacity and \textit{do not become bottlenecks}. We use past works \cite{switchboard:sigcomm23, saving:conext24, xron:sigcomm23} for P1.


\subsection{Findings from \teams production}
\label{sec:back:workload}

\begin{table}
\centering 
\caption{Distribution of number of participants in individual calls. Exact numbers not shown due to business sensitivity.}
\label{tab:back:num}
{
\small
\begin{tabular} {|c|c|c|c|c|}
\hline
Percentiles & P10 & P50 & P90 & P95\\
\hline
\#Participants & 2-3 & 3-5 & 8-13 & 11-15 \\
\hline
\end{tabular}
}
\end{table}

\begin{figure*}
\centering
    \begin{minipage}{.45\textwidth}
        \centering
        \includegraphics[width=0.95\textwidth, page=1]{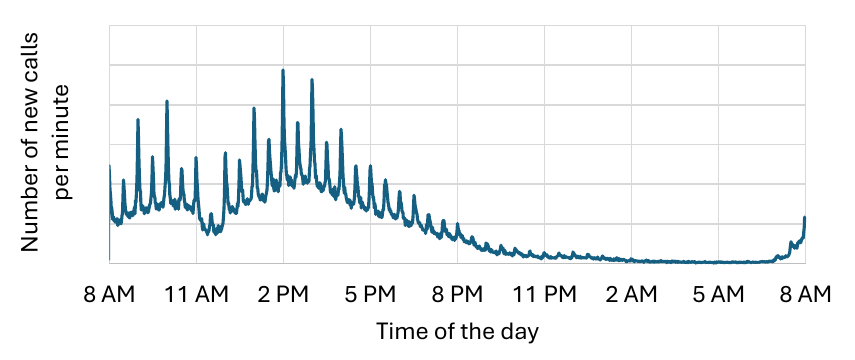}
        \caption{Calls arrive in bursts around 30 and 60 minute marks. Time shown in local timezone. The y-axis values are not shown due to business sensitivity.}
        \label{fig:measure:arrival}
    \end{minipage}%
    \hspace{0.2cm}
    \begin{minipage}{0.45\textwidth}
        \centering
        \includegraphics[width=0.95\textwidth, page=2]{figs/measurement-results-paper.pdf}
        \caption{CPU utilization (max, P95, P50 and average) across all MPs in the same DC. P50 and average lines mostly overlap. The y-axis values are not shown due to business sensitivity.}
        \label{fig:measure:cpu}
    \end{minipage}%
\end{figure*}


Large-scale conferencing services such as \teams have unique characteristics in their workload that differ from other workloads such as web-servers or caches or analytic jobs\cite{insidefb:sigcomm15}. For example, the calls in \teams are bursty and arrive at 30- and 60- minute mark, and last for minutes to hours with call sizes changing as participants join/leave. To illustrate the unique characteristics, we observe all calls for 24 hours in a randomly chosen DC (our findings apply to all other DCs too). We detail our findings below.

\textbf{(Finding-1) Call sizes differ significantly.} Table \ref{tab:back:num} shows the distribution of max. number of participants ($N$) in individual calls. It can be seen that there is a stark difference in $N$ across calls where the P50 is between 3-5 and P90 is between 8-13 (exact numbers not shown due to business sensitivity).  This disparity in the distribution of $N$ poses the first challenge in assigning the MPs. The \teams controller needs to make assignments when the \textit{first} participant joins the call. When calls start, all calls look the same -- only 1 participant has joined. However, over a period of time, different calls can grow to radically different sizes. The controller does not know the future CPU demand by the calls and can lead to poor load balancing and hot MPs.

\textbf{(Finding-2) Bursty traffic pattern.} Fig.\ref{fig:measure:arrival} shows the number of calls started every minute. It can be seen that substantial calls started at 30 and 60 minute marks. A drop at around 11:30AM is due to lunch hour. Such bursty behavior poses two challenges: (a) as mentioned above, the size of calls could differ significantly. In addition to dealing with uncertainty of one call, \teams controller needs to deal with uncertainty of multiple calls simultaneously. This exacerbates the challenge mentioned above. (b) the controller needs to scale to support burst of calls near the 30 and 60 minute mark; while the controller is relatively idle at other times.

\textbf{(Finding-3) High imbalance in CPU utilization.} Fig.\ref{fig:measure:cpu} shows the imbalance in CPU utilization across MP servers in the same DC. First, we measure the CPU utilization on all MPs every minute. We then measure the max, P95 (95$^{th}$ percentile), P50 and average across all MPs. Fig.\ref{fig:measure:cpu} shows the CPU utilization for 24-hour period. It can be seen that max and P95 are substantially higher than P50 and average even at the busiest time -- the max. CPU utilization is 1.63$\times$ P50 at the busiest time. 

The key reason for such an imbalance is the \textbf{variability in call sizes} due to which the controller does not know the call size (number of participants and media streams) ahead of time. As calls progress (new participants join and/or start video streams), the CPU utilization on the MPs change. Even if two MPs have the same number of calls, due to the differences in the call sizes, one MP can have radically higher CPU utilization than other MP. \rohan{The load balancer detects the high CPU utilization and uses the other MP nodes for the subsequent calls that just move the hotspots as explained next. Note that \teams does not move active calls frequently to avoid the undesired effects on the user experience}  

\begin{figure}[t]
\centering
\includegraphics[width=0.45\textwidth, page=1]{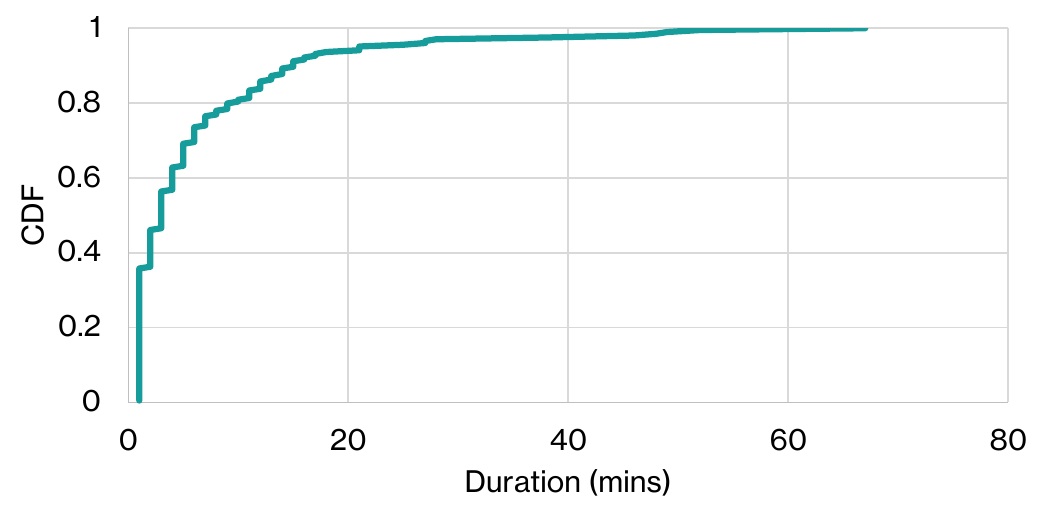}
\caption{CDF of duration for which individual MPs ran hottest.}
\vspace{-0.1in}
\label{fig:measure:hotcpu}
\end{figure}

\textbf{(Finding-4) Hot nodes run for short duration and mostly don't repeat.} In this experiment, we record the MP server with max. CPU utilization in every minute from production DC for 24 hours. Fig.\ref{fig:measure:hotcpu} shows the CDF of the duration for which individual MPs were hottest. We make two observations: (a) mostly, once an MP server becomes hottest, it remains so for a few minutes. As shown in Fig.\ref{fig:measure:hotcpu}, the P50 and P90 of the duration is 3 and 15 mins, (b) mostly, the hottest MP rarely returns as hottest MP again. This is because, once the MP becomes hot, other (colder) MPs get picked for assigning the next new calls. As a result, the CPU utilization on the hot MPs mostly does not increase; instead, it comes down as its calls end. As next new calls are placed on other MPs, their CPU utilization increases leading them to become hot. There are a few exceptions. We observed that one MP was hottest cumulatively for roughly 67 minutes. However, such an MP was the hottest for max. 12 minutes in a stretch. It was just that more calls were assigned to it once it cooled down a bit, making it hot again. 

\textbf{(Finding-5) Call migration is infrequent.} One solution to reduce the peak CPU utilization is to migrate the calls midway. However, it requires call state transfer from one MP to another. Additionally, the \teams clients are to be updated with the new MP. This takes time and at times results in call glitches or (worse) call drops. We try to minimize the call migrations. 

\textbf{(Finding-6) Heterogeneous hardware.} Our VMs run on hardware with different CPU vendors and CPU generations. Thus, the same call results in different CPU utilization on different VMs potentially leading to hot MPs when calls get placed on slower CPU. However, such impact is predictable depending on the CPU SKU.

\textbf{Stuck between rock and hard place.} The CPU imbalance translates into poor user experience or higher costs (requiring more MPs). To keep the MPs cold, we need significant overprovisioning as max. CPU utilization is significantly higher than P50 and average. Such high overprovisioning balloons the  costs for running \teams, which is not desired. Alternatively, without overprovisioning, we risk running some of the MPs hot, degrading user experience. \S\ref{sec:problem} sheds more light on this issue. 

\section{Limitations of existing LB algorithms}
\label{sec:problem}

As mentioned in \S\ref{sec:back:workload}, there is a significant CPU imbalance across MP servers in the same DC. Note that, such an imbalance is not only due to the algorithm used in \teams but also occurs across the rich body of work on load balancing (or packing) traffic across servers including algorithms such as round robin, least loaded server etc. Since we cannot evaluate these other algorithms on \textit{live} traffic, therefore we simulate their performance using a trace as described below.

\textbf{Trace.} We collected a trace of anonymized calls on a typical workday (24 hours) from a randomly chosen DC. The trace contains the time of arrival for each participant in a call. The trace also records the start and end time for each media type (audio, video and screen share). Our results apply to all DCs for \teams.

\subsection{Simulations}
\label{sec:problem:sim}

\textbf{Simulator to replay call trace.} The simulator replays the call trace of 24 hours under different scenarios. With the simulator, we can change the MP assignment algorithm, cluster size and migration algorithm. 

The simulator calculates the CPU utilization for each MP at each minute. Recall that the CPU utilization is a function of total network traffic handled, which in turn is a function of number of participants and their media type (\S\ref{sec:back:mp}). Thus, we calculate CPU utilization using its participants and media type (and quality) captured in the trace. The CPU utilization for an MP is simply the sum of CPU utilization for all its calls. The simulator assumes homogeneous compute, and does not account heterogeneity in hardware. The CPU calculation has high fidelity and calculates CPU utilization of a call within 8\% of real CPU utilization of a call.

The simulator then calculates the maximum (max), minimum (min), P95, P50 and average CPU utilization every minute across all MPs. Additionally, every minute, it also calculates the number of hot MPs, i.e., the MPs with at least 75\% CPU utilization. This was a threshold obtained from the \teams production team. For lower CPU utilization, there are small queues formed at CPU. However, queue lengths accelerate when the CPU utilization is high as CPU cannot keep up with the packets arriving; potentially leading to (undesired) packet drops. Thus, it is important to keep a lid on CPU utilization. Prior works have made similar observations\cite{knapsacklb:conext25}. Lastly, it calculates the number of calls and participants on such hot MPs.


\textbf{Baseline algorithms.} Due to business sensitivity, we do not describe the exact algorithm used for MP assignment in \teams, although it is akin to load balancer algorithms\cite{nginx:web, lbalgo:web}. Specifically, our algorithm is closest to LLR detailed below. 

We outline three well established and state-of-art baselines that assign the MP to a call \textit{at-scale} when the first participant on that call arrives. (a) \textbf{Round robin (RR).} This policy simply rotates calls across individual MPs. We choose such a baseline as it's simple to implement and balances calls equally among MP servers. (b) \textbf{Least load (LL).} RR is oblivious to the load on the MP. LL selects the MP with least load at the time of assignment reducing the risk of MPs becoming hot, (c) \textbf{Least load random (LLR).} This strategy first selects $K$ least loaded MPs and then selects one MP out of $K$ uniformly at random. This addresses the limitation of LL that it may end up assigning many calls to the same MP. For example, when we assign an MP when the call starts, the increase in CPU utilization due to that call at start of the call is small as only a few participants have joined. Thus, the same MP may continue as the least loaded MP and more calls can get added to the same MP. As calls progress, the CPU utilization will increase and risk that MP becoming hot. LLR addresses this limitation by spreading new calls across $K$ least loaded MPs at that time. We set $K$ = 5 for experiments.

\textbf{Migration strategy.} The above algorithms only select MP \textit{once} when the calls start. However, as the calls progress (more participants join and/or media changes), the CPU utilization on some MPs can increase causing them to become hot. The above algorithms will simply not assign the next calls to such hot MPs, but will not make any adjustments on their own to reduce the load on hot MPs. Thus, in addition to these algorithms, we \textit{migrate} a subset of calls from hot MPs to other (cold) MPs to cool down such MPs. We consider a simple yet scalable migration policy detailed below. 

We first divide all MPs into hot and cold MPs based on their CPU utilization. MPs are hot if their CPU utilization $\geq T$ ($T$ = 75\%) while the rest are cold. We consider all hot MPs in decreasing order of their CPU utilization. For every hot MP, we first select the calls in random order for potential migration based on their CPU utilization. Note that if we select the calls in descending order of CPU utilization (elephant calls first), we fail to migrate the calls as potential cold nodes may not have capacity to host such elephant calls. On the other hand, if we select calls in ascending order (mice calls first), we end up migrating many calls. To strike a balance between these choices, we choose to randomly select the calls. For each call, a target MP is selected using "First Fit", where we assign it to the first MP that has capacity to host such a call. We stop migrating calls from one hot MP when some of its calls are migrated and the CPU utilization on that MP falls below $T$. The migration policy is independent of the algorithm used to assign the MP for individual calls and thus works with all above baseline algorithms. We run migration every 2 mins.

\textbf{Metrics of interest.} For each assignment algorithm (with and without migration), the metrics of interest are: (a) imbalance in CPU utilization, (b) user experience driven by number of hot MPs, and the number of participants and calls on the hot MPs. \textit{Hot MPs are those with CPU utilization exceeding 75\%}, (c) cost of service (number of MPs required), (d) number of migrations. 

\textbf{Cluster configuration.} We consider two cluster sizes -- 3600 and 3000 MPs. We choose 3600 where average CPU usage is similar to that in production. As we are not reshaping the calls, the average is same across all assignment algorithms. One way to bump up the average CPU utilization is to reduce number of MPs. Thus, we consider  cluster of 3000 MPs too.

\subsection{Limitations}
\label{sec:back:limit}

\begin{table}
\centering 
\caption{Ratio of max to average CPU utilization at busiest time. In contrast, such a ratio in production is 1.63.}
\label{tab:problem:cpu}
{
\small
\begin{tabular} {|c|c|c|c|c|}
\hline
Cluster size & Migration & RR & LL & LLR \\
\hline
\multirow{ 2}{*}{3600} & No & 2.62 & 1.88 & 1.82 \\
 & Yes & 1.36 & 1.36 & 1.32 \\
\hline
\multirow{ 2}{*}{3000} & No & 2.14 & 1.85 & 1.63 \\
 & Yes & 1.32 & 1.38 & 1.27 \\
\hline
\end{tabular}
}
\end{table}

\begin{figure*}[t]
\subfigure[Number of participants on hot MPs]
{
\includegraphics[width = 0.33\textwidth, page=2]{figs/motivation-2.pdf}
\label{fig:problem:calls}
}
\hspace{-0.2cm}
\subfigure[Max. CPU utilization. 1 indicates 100\%.]
{
\includegraphics[width = 0.33\textwidth, page=1]{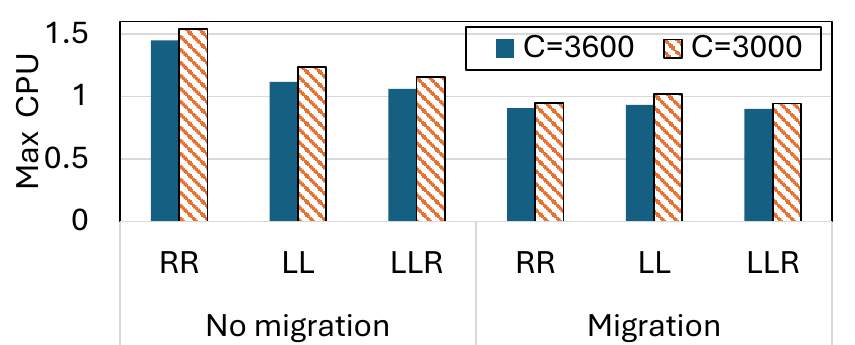}
\label{fig:problem:cpucluster}
}
\hspace{-0.5cm}
\subfigure[Number of calls migrated]
{
\includegraphics[width = 0.33\textwidth, page=3]{figs/motivation-2.pdf}
\label{fig:problem:migration}
}
\caption{Performance of different baseline algorithms at different cluster sizes (C).}
\vspace{-0.3cm}
\protect\label{fig:back:clustersize}
\end{figure*}

\textbf{CPU imbalance.} First, we show the CPU imbalance due to the 3 baseline strategies with and without migration. Note that, migration does not kick in frequently in the production as max. CPU utilization is under control due to over-provisioning. Table~\ref{tab:problem:cpu} shows the ratio of max to average CPU utilization (denoted by $M$) across all MPs at the \textit{busiest} time when MP usage is highest. A higher value of $M$ denotes greater imbalance. \textbf{(a) Without migration.} It can be seen that all the algorithms end up with high $M$. RR performs worse as it is oblivious to the CPU utilization. LLR performs better than other algorithms as it considers CPU utilization and addresses limitations of LL. $M$ is high irrespective of the cluster size. \textbf{(b) With migration.} All algorithms have similar values for $M$ when using migration. Note that, all the baseline algorithms differ in initial placement. With migration, the flaws of initial placement are mitigated as the migration policy migrates the calls to reduce the load on the hot MPs. As the migration policy is common for all baseline algorithms, we observe similar performance for all such baseline algorithms when using migration.

\textbf{Number of participants on the hot nodes.} Next, we show the number of participants on the hot nodes (denoted by $N$) in Fig.\ref{fig:problem:calls}.  This metric is important as it shows the number of participants potentially suffering from poor user experience. For each minute of the day, we calculate number of participants on the hot MPs, and $N$ denotes the total number of participants on hot MPs across 24 hours. Note that if an MP is running hot, \textit{all} its participants suffer from poor experience. For a cluster size of 3600, LLR results in fewer hot nodes and consequently smaller $N$. Migration also leads to significant reduction in $N$. Using LLR and migration, the $N$ for a cluster size of 3600 is just 5600. However, as we reduce the cluster size to 3000 MPs, we see a significant jump in $N$. For LLR with migration, there is roughly 12$\times$ increase in $N$. Similarly, peak number of participants on hot MPs relative to total active participants at a given time jumps by 8$\times$, which is not acceptable.


\textbf{Max. CPU utilization.} Next, we evaluate the algorithms in terms of max. CPU utilization. To do so, for each minute of a day, we calculate the max CPU utilization across all MPs. We then take max. of such  values across a day. We consider P95, minimum  CPU utilization, and show them in Fig.\ref{fig:eval:cpu} (\S\ref{sec:eval}). CPU utilization is important as it shows the extent of poor performance -- higher the CPU utilization on hot nodes, higher the extent of poor performance. As shown in Fig.\ref{fig:problem:cpucluster}, the max. CPU utilization increases substantially when we reduce the cluster size from 3600 to 3000 (for all algorithms with/without migration) indicating that cutting down the cost may degrade user experience. CPU utilization above 100\% in some cases is an artifact of simulations. In real cases, the CPU utilization is bound to 100\%. CPU utilization above 100\% indicates higher packet loss. 

\textbf{Migrations.} The previous experiments show that migration helps substantially to cool down the MPs -- both in terms of max. CPU and number of participants on the hot MPs. However, migration may cause temporal glitches as we migrate calls from one MP to another, and put a strain on the controller to ensure seamless migration. In this experiment, we measure the number of migrations for 24 hour period. Fig.\ref{fig:problem:migration} shows LLR results in the least migrations, but migrations increase substantially (9$\times$) as we reduce MPs from 3600 to 3000.

\textbf{Summary.} Our analysis shows that the existing policies of using RR, LL and LLR fall short. Using larger cluster size, the user experience is good as very few calls end up running on hot MPs. However, it suffers from lower cluster utilization (and hence higher costs). On the other hand, if the cluster utilization is improved by cutting down the number of MPs, users may get poor user experience as the number of calls on hot instances increases. Call migration helps but such a policy results in high number of migrations.

\section{\name}
\label{sec:name}

In this paper, we present \name (Packers and Movers), which aims to \emph{pack} the calls across MPs with the goal of  providing a better user experience (fewer hotspots) for a given set of MPs in \teams. \name runs independently in each DC and manages calls and MPs only in that DC. While in this context, packing and load-balancing are synonymous, \name departs from classic load balancing algorithms in two crucial ways: a) the initial assignment of calls to MPs, when the first participant joins, is adaptive and based on historical patterns unlike the baselines (RR, LL, LLR) of the previous section which only use the current CPU utilization and b) unlike classic load balancing algorithms where MP hosting the call is fixed after the initial assignment, \name subsequently \textit{moves} the call to a different MP (in the same DC) when the call size is converged. 

\subsection{Key ideas}
\label{sec:key}

We now summarize our key ideas. As mentioned in the previous section, existing algorithms use \textit{current} CPU utilization when assigning calls to MP nodes. Unfortunately such algorithms fall short because call size (and consequently CPU utilization) is not known apriori and fluctuates over time. However we observed that \teams has rich, yet hitherto overlooked data on \emph{call history} that can be used to estimate the call sizes, which serves as an indicator of the peak CPU utilization. We can pre-allocate CPU based on the peak to reduce the chances of MPs becoming hot later. This is captured in the first two key ideas below. The third idea below captures how we react to incidents of hot MPs and migrate the calls away from those servers. 



\textbf{(K1) Leveraging call history for recurring calls.} We observe that a large chunks of our calls have history. Many calls are a part of a \textit{recurring call series} (e.g., calls that repeat daily or weekly) and observe similar number of participants. We leverage the history to predict the peak call CPU requirement, and assign an MP that can hold the peak CPU requirement of that call to minimize the chances of MPs getting hot. We found that roughly 40-60\% (actual number not mentioned due to sensitivity of the information) calls are recurring. All recurring calls within a given call series have the same \textit{call ID}, which helps identify the history of previous calls of the same call series. 

\textbf{(K2) Estimating CPU utilization for non-recurring calls.} Our second key idea is to predict the peak CPU requirement for calls with no history (non-recurring calls). We do so by building a model that trains using history of non-recurring calls, and predicts the peak CPU utilization of a call based on its age, current number of participants and media type. Subsequently, it helps us \textit{migrate} the call by taking into account the (future) peak CPU utilization of the call. 

\textbf{(K3) Call migration.} Due to intrinsic nature of the usage of the call, the above key ideas do not eliminate the instances of hot MPs and migrations. For example, even for recurring calls, there is some variance in number of participants and media type (audio, video etc.). For non-recurring calls, the model is not 100\% accurate. Our last key idea is to \textit{wait} for some time once the call starts so that its call size (and its CPU utilization) stabilizes, and then \textit{migrate} the call so as to reduce hotspots. The key challenge though is to decide which calls are to be migrated and also their target MP. We develop an Mixed Integer Program (MIP) to do the migration.

\subsection{Architecture}
\label{sec:key:arch}

\begin{figure}[t] 
\centering
\includegraphics[width=0.42\textwidth, page=1]{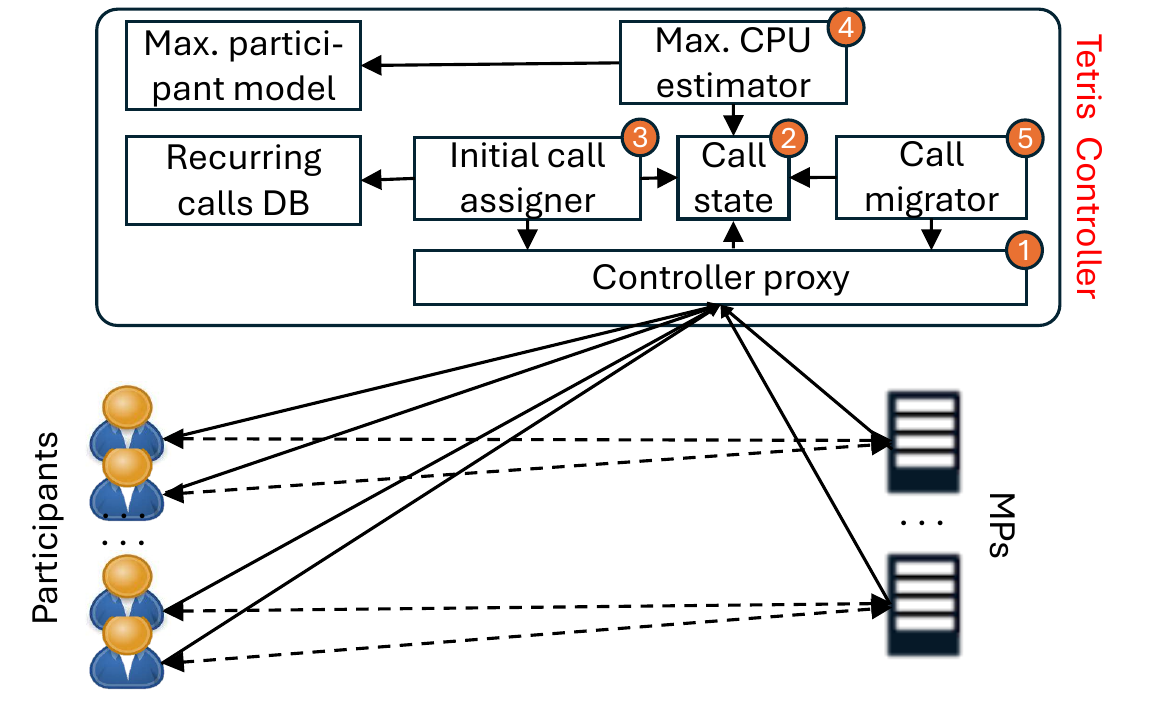}
\caption{Architecture of \name. Solid lines are control signals while the dotted lines denote call data traffic.}
\vspace{-0.1in}
\label{fig:key:arch}
\end{figure}

Fig.\ref{fig:key:arch} shows the architecture of the \name. It has three components. Participants and MP servers are unchanged from \teams. 

\ankur{Seems a little weird to say PnM controller has 3 components and one of them is PnM controller.}
\rohan{changed, thanks.}

\textbf{Participants (\name clients).} When the first participant of the call joins, the controller first assigns the MP server that will host the call.  For the subsequent participants on the same call, the controller simply relays the MP server assigned. This way, all the participants of the call are assigned to the same MP server. During the call, if participants change the media type (e.g., video or screen-share), or change the quality of the media type (e.g., video changed from 720p to 1080p), the participants relay that information to the controller.

\ankur{What does the previous sentence mean?}
\rohan{removed it}

\textbf{MP servers.} The MP is detailed in \S\ref{sec:back:mp}. MP servers receive, mix and re-transmit the media streams to all participants of the same call (dotted lines in Fig.\ref{fig:key:arch}). Each MP server hosts multiple concurrent calls. When the MP becomes hot, it cannot keep up with the rate at which network packets are received, and starts dropping network packets once the queues are full. As a result, potentially \textit{all} calls on that MP suffer. A temporary fix to overloading MP is that the MP instructs participants to send videos at lower quality. However, in \name, we try to assign (or migrate) the calls so that we meet the best quality MP has seen so far.

\textbf{\name controller.} \name controller performs three actions: (a) assign the MP server when the call starts, (b) periodically recompute the max. CPU utilization of calls based on age and current number of participants (Key idea K2) and, (c) migrate the calls to cool down hot MPs (Key idea K3). To do so, it has 5 modules that we detail below:

(1) \textit{Controller proxy.} This module communicates with all participants and MPs. The participants push the changes in media type and quality to \name controller through this module. Similarly, MPs push their CPU utilization every minute. Controller proxy creates a queue where all these events are pushed.

(2) \textit{Call state.} This module contains all the information about the call including number of participants, media type and quality for each participant. Using this information, it also computes the expected CPU utilization for every call. As mentioned in \S\ref{sec:design}, we use LLR in conjunction with \textit{expected (peak)} CPU utilization that preallocates CPU based on future growth of the call. All the remaining modules read and update the state in this module.

(3) \textit{Initial call assigner.} For each new call, it first checks if the call is part of any recurring series. If so, it pulls the number of participants and media type seen previously for the recurring series of such a call. It assigns the call using such history as mentioned in \S\ref{sec:design:recurring}. \name stores max. participants as well as the media type for recurring calls in DB (Recurring calls DB in Fig.\ref{fig:key:arch}). For non-recurring calls, it uses LLR to assign the MP as detailed in \S\ref{sec:design:ankur}.

(4) \textit{Max. CPU estimator.} For non-recurring calls, we use key idea K2 to estimate the peak CPU utilization of a call. We build a model that uses previous 7 days of data and we refresh it every 24 hours.  The model is then stored as a lookup table (Max. participant model in Fig.\ref{fig:key:arch}) where the key is a tuple of age and number of current participants and value is the max. number of participants. This module runs every minute and refreshes the max. CPU utilization for every non-recurring call (more details in \S\ref{sec:design:ankur}). 

(5) \textit{Call migrator.} Once the calls progress, the MPs can turn hot as prediction is not 100\%. In such cases, we migrate the calls (call migrator in Fig.\ref{fig:key:arch}). We run migration every two minutes. The migrator not only considers calls on current hot MPs but also considers calls on MPs that are near to getting hot. This helps it address the calls on MPs that might become hot during 2 minutes. 

\ankur{We should explicitly define what "hot MP" means somewhere.}
\rohan{done.}

\section{Initial Call Assignment}
\label{sec:design}


\subsection{Initial call assignment for recurring calls}
\label{sec:design:recurring}

\begin{figure*}[t]
  \centering
  \includegraphics[width=0.95\textwidth, page=1]{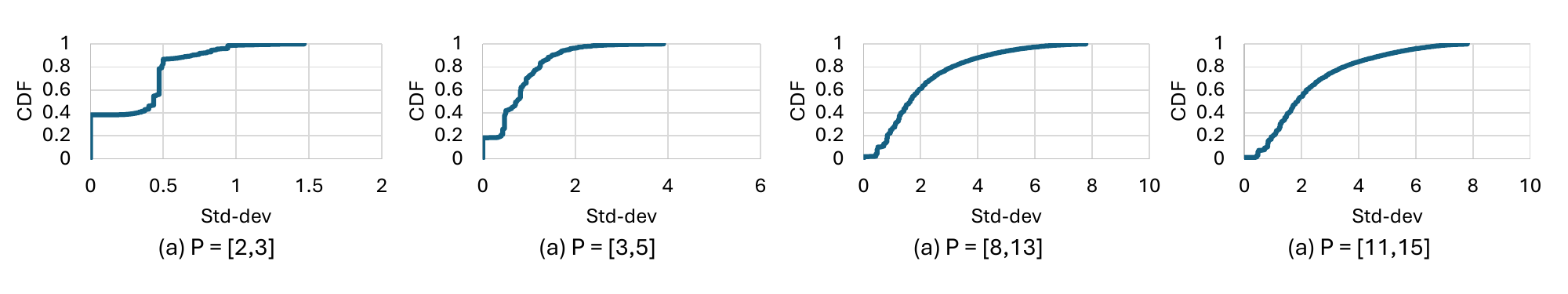}
  \caption{Standard deviation of number of participants across individual call series for  participant sizes (P) from Table \ref{tab:back:num}.}
  \label{fig:design:stddev}
\end{figure*}

Roughly 40-60\% calls in \teams are a part of recurring series (\S\ref{sec:key}). We observe that max. number of participants and media type have good predictability for calls of such series. Fig.\ref{fig:design:stddev} shows the CDF of standard deviation of max. participants across individual call series. We consider all calls within 4 weeks and call series where we have minimum 4 calls in individual call series. Across all calls (not shown), for 20\% call series, max. number of participants do not change while for roughly 65\% call series, the standard deviation is $\leq$ 1. This shows that we can predict number of participants fairly accurately for many call series. We see similar trends for media types too. 

\ankur{I think just keeping the CDF might be enough here. We don't need both plots}
\rohan{done}

\textbf{Predicting peak number of participants.} We use the history for the recurring calls to estimate the peak number of participants and media types for the new call. We use simple weighted moving average that we found to work well. For a call, we first find if it's a part of a recurring series with at least 4 previous occurrences. We estimate number of participants for the new call as: 0.5$\times$ p$_{0}$ + 0.25$\times$ p$_{1}$ + 0.125$\times$ p$_{2}$ + 0.125$\times$ p$_{3+}$. p$_{0}$ to p$_{2}$ indicate peak number of participants from the latest to the third-latest occurrence. P$_{3+}$ is average of number of participants for all other older occurrences. Such a design is simple to implement, and helps \name controller do assignments quickly at scale. We calculate the media type similarly. Lastly, based on the estimated number of participants and media type, we calculate expected peak CPU utilization (using module in \S\ref{sec:problem}). Recurring calls with fewer occurrences are treated as non-recurring calls.

\textbf{Addressing heterogeneity.} We calculate the peak CPU above for the most common CPU SKU. As mentioned in \S\ref{sec:back:workload}, \teams runs on MPs from different CPU SKUs. Offline, we profile the performance of MP on all SKUs for a mixed workload of calls and measure the ratio of performance across different SKUs. Online, we use the ratio to calculate CPU utilization on different CPU SKUs.    

\begin{table}
\centering 
\caption{Distribution of call joiner duration (difference between the times when last participant and first participant joining the calls. This is over millions of calls in 24 hours.}
\label{tab:design:duration}
{
\small
\begin{tabular} {|c|c|c|c|c|}
\hline
P50 & P75 & P95 & P99\\
\hline
12 sec & 46 sec & 293 sec & 550 sec \\
\hline
\end{tabular}
}
\end{table}

\textbf{Assigning the MP for recurring calls.} Once we calculate the (expected) peak CPU utilization in previous step, we do LLR using peak CPU utilization. Most of the calls start at 30- and 60- minutes mark (Fig.\ref{fig:measure:arrival}). Additionally, we observe that most of the participants on the calls join in first few minutes and stay for the duration of the call (Table \ref{tab:design:duration}). As a result, the peak CPU utilization of an MP is the sum of peak CPU utilization of all its calls (including recurring and non recurring calls). Note, the peak CPU utilization for non-recurring calls is explained in \S\ref{sec:design:ankur}. The LLR used in baselines (\S\ref{sec:problem}) use LLR with \textit{current} CPU utilization while our estimate of peak CPU utilization pre-allocates the CPU for the \emph{future growth} in call size. We calculate 5 MPs with the least expected CPU utilization and select one MP uniformly at random. Note that, we do not waste CPU cycles by provisioning for peak as peak is reached quickly as calls start (Table \ref{tab:design:duration}).

\vspace{-0.2in}

\subsubsection{Discussion}

\textbf{Call packing.} As we predict the CPU utilization for the recurring calls, there is an opportunity to do \textit{best fit} to improve CPU utilization on the MPs. However, we found that best fit performs worse than doing LLR as we still need to place  non-recurring calls alongside recurring calls on the same MPs. The increase in the CPU utilization due to non-recurring calls and inaccuracy in prediction for recurring calls sends MPs hot. Thus, we resort to LLR.

\textbf{Call duration data.} Secondly, we currently do not use the call duration information from historical data. Such data can help us do better call packing. We leave it to the future work.

\textbf{Calendar data.} Lastly, we also do not use the calendar data. This data is not available due to the privacy requirements. However, as shown in \S\ref{sec:eval:recurring} such information could be useful. That said, we leave using the calendar data to the future work.

\subsection{Initial assignment for non-recurring calls}
\label{sec:design:ankur}



\textbf{Predicting peak number of participants for non-recurring calls.} Let $N_{\max}(n,t)$ be our estimate of the max. participants of a non-recurring call with $n$ participants at time $t$ (from the start of the call). Let $p_t(x)$ be the number of participants at time $t$ for the call with meeting ID = $x$. Let $c(x)$ be the max. number of participants in the call with meeting ID = $x$. Then our estimate for $N_{\max}(n,t)$ is:
\begin{align}
    N_{\max}(n,t) = \frac{\sum_{m \geq n}w_{m}m}{\sum_{m \geq n}w_{m}} \label{eq:maxest}
\end{align}
where $w_m = \sum_{x} \mathbbm{1}\{p_t(x) \leq n, c(x) \geq m\}$.

To understand the rationale behind the above heuristic observe that $N_{\max}(n,t) \geq n$ by definition. We focus on \emph{only} those calls which had fewer than $n$ participants at time $t$ and then eventually attained a maximum number of participants that was larger than $n$ because:

(1) Calls with more than $n$ participants at time $t$ are assumed to be growing faster than the current call and therefore the current call may not be able to keep up with their growth rate or maximum number of participants.

(2) Calls with maximum participants less than $n$ are too small and the current call has already exceeded their size.

Therefore, we estimate $N_{\max}(n,t)$ as the weighted average of all values $m \geq n$, where each $m$ is weighted by the fraction of calls with fewer than $n$ participants at time $t$ but eventually attained a max. number of participants  $\geq m$. This weight serves as our estimate of the probability that calls with $n$ or fewer participants at time $t$ end up having at least $m$ participants. We update our estimates of $N_{\max}(n,t)$ every minute.

Lastly, we multiply peak number of participants with average media rate to estimate the peak CPU utilization. We adjust the CPU utilization depending on the CPU SKU (\S\ref{sec:design:recurring}). Note that the estimated peak CPU utilization for non-recurring calls varies over time as the estimate $N_{\max}(n,t)$ in \eqref{eq:maxest} varies over time. Hence, the expected CPU utilization of the host MP also needs to be adjusted so that MPs with high expected CPU utilization are not considered for assigning new calls.


\textbf{Assigning the MP for non-recurring calls.} For non-recurring calls, the "Max CPU estimator" module (module-4 in Fig.\ref{fig:key:arch}) recomputes peak CPU utilization of calls and MPs every minute. Once again, we use LLR while considering the expected peak CPU utilization on each MP.  

\section{Call migration}
\label{sec:migration}

\begin{table}[t] 
\centering 
\caption{Notations used in the MIP.}
\label{tab:algo:notation}
{
\small
\begin{tabular} {|c|c|}
\hline
\textbf{Notation} & \textbf{Explanation}\\
\hline
\hline
\multicolumn{2}{|c|}{Input} \\
\hline
$M,C$ & Set of MPs and calls \\
\hline
$P_{c}$ & Estimated max. CPU usage of c-th call\\
\hline
$B_{m}$ & CPU usage on m-th MP due to stationary calls\\
\hline
$R_{m}$ & Ratio of performance for CPU SKU of m-th MP \\
\hline
$Cap_{m}$ & CPU capacity on m-th MP\\
\hline
$O_{m,c}$ & Set if c-th call is placed on m-th MP originally\\
\hline
$L$ & Limit on number of migrations\\
\hline
\hline 
(output) $X_{m,c}$ & set if c-th call is assigned to m-th MP \\
(output) $y$ & max. CPU usage across all MPs \\
\hline
\end{tabular}
}
\vspace{-0.1in}
\end{table}


Previously, we detailed \name for initial assignment of calls. Our initial assignment may lead to hot MPs as: (a)  prediction is not always accurate, (b) non-recurring calls whose future demand is not known at the time of assigning MPs. Consequently, we \textit{wait} for the call sizes to stabilize, and \textit{migrate} calls from hot to cold MPs.

The goal of call migration is to decide what calls to migrate and their target MP so that the CPU utilization on hot MPs comes down without creating new hot MPs. 

A straw-man's approach could be to use the simple migration strategy detailed in \S\ref{sec:problem:sim}. However, such a greedy choice is not optimal as it does not use the (holistic) knowledge across calls and MPs. Our insight is that migration problem could be modeled as a bin packing optimization problem. To solve it, we formulate it as a Mixed-Integer-Problem (MIP) tailored for migration, and solve it using off-the-shelf MIP solver (COIN-OR\cite{coinor:web}). However, solving it in a timely manner is challenging that we address in the next section.

The notations for MIP are shown in Table~\ref{tab:algo:notation}, while Fig.\ref{fig:algo:lp} shows the MIP formulation. The objective of the MIP is to minimize the maximum CPU utilization across all MPs. MIP variables are: (a) $X_{m,c}$ (binary) -- set when the c-th call is assigned to m-th MP. (b) $y$ (between 0-100\%) -- max. CPU utilization across all MPs.

We choose such an objective to minimize the calls on hot MPs. Note that even though we consider calls that have been running for a while, the CPU usage of the calls can still change as participants join/leave or due to changes in media. By minimizing the max. CPU utilization, we keep more room for such cases. We give more details on calls considered for migration in the next section. In \name, we do not minimize the number of MPs as it is complicated by other factors such as failure resiliency\cite{switchboard:sigcomm23}. Thus \name runs with given set of MPs, and packs calls for better performance.

\begin{figure}[t] 
{\small
\fbox{
  \begin{minipage}{0.45\textwidth}
    \textbf{MIP Variable:} $X_{m,c}$, $y$\\ 
    \textbf{Objective:} Minimize y\\
    \textbf{Constraints:}\\
    Only one MP assigned to each call: $\forall c \in C, \displaystyle\sum_{m \in M} X_{m,c} = 1$ \hfill (a)\\
    CPU under limit: $  \forall m \in M, \displaystyle\sum_{c \in C} X_{m,c} \cdot P_c \cdot R_m + B_{m} \leq Cap_{m}$ \hfill (b)\\
    Number of migrations: $\displaystyle\sum_{m \in M}\displaystyle\sum_{c \in C} (O_{m,c} - X_{m,c}) \cdot O_{m,c} \leq L$ \hfill (c) \\
    Expressing y: $\forall m \in M, y \geq \displaystyle\sum_{c \in C} X_{m,c} \cdot P_c \cdot R_m + B_{m}$ \hfill (d)
  \end{minipage}
}
}
\caption{MIP formulation.}
\vspace{-0.15in}
\protect\label{fig:algo:lp}
\end{figure}

There are four constraints in Fig.\ref{fig:algo:lp}. (a) Each call is assigned to exactly one MP. (b) The total CPU utilization on each MP is under the threshold ($Cap_{m}$). Note that we do not consider all calls to migrate (more details in \S\ref{sec:lp:input}) to improve scalability. E.g., calls on the (cold) target MPs are not considered for migration. $B_{m}$ denotes the CPU utilization on m-th MP due to such stationary calls. (c) Number of migrations is under the threshold ($L$). (d) We express $y$ as max. CPU across all MPs.

Constraint (c) addresses the migration limits. When a call is moved to a new MP, it may get counted twice as it is moved away from old MP and added to new MP. However, (c) ensures we only count it once (on old MP). Such a condition only needs to check when $O_{m,c}$ is set. Note, $O_{m,c}$ is an input and not MIP variable.

\subsection{Speeding up MIP}
\label{sec:lp:input}

The goal on the MIP is to \textit{effectively} and \textit{quickly} migrate the calls. The execution time of the MIP is important as a high execution time means that the hot MPs will remain hot for longer. The total number of MIP variables are $|M| \times |C|$, which could be 10s of millions at busy times. Ideally, we could consider all calls in $C$ and all MPs in $M$ to optimally calculate the migration plan. However, this has two shortcomings: (a) the total number of MIP variables simply overload the MIP solvers where we cannot solve MIP quickly, and (b) shuffling of calls between \textit{cold} MPs may have minor improvements in objective. To reduce the MIP execution time, we limit calls in $C$ and target MPs in $M$ as detailed below.

We have four optimizations to speed up MIP:
(1) The input to the MIP includes $C$ -- calls to be migrated. \textit{We do not consider mice calls on the hot MPs} as candidate calls to be migrated. Candidate calls have a peak CPU utilization exceeding $E$ (set to 2\%). This design choice reduces the number of calls in $C$ that drastically reduces the MIP execution time while still packing the calls effectively. We also consider non-mice calls on MPs with peak utilization near threshold for hot MPs (95\% of $Cap_{m}$).

(2) The input also includes $M$ -- set of MPs to host the calls. This set of MPs includes all hot MPs as some calls may remain on those MPs. Additionally, to reduce the MIP execution time, we only consider a subset of cold MPs where the calls can be migrated. We decide the cold MPs as follows: 

The total CPU (denoted by $T1$) exceeding threshold $Cap_{m}$ on hot MPs is $T1 = \displaystyle\sum_{m \in HotMP}\displaystyle\sum_{c \in C} P_{c} \cdot R_{m} \cdot O_{m,c} - Cap_{m} \cdot |HotMP|$. We select the target MPs with cumulative available capacity of 5 $\times$ T1. We sort the cold MPs in descending order based on the peak CPU utilization, and pick the MPs from the top till the total available CPU exceeds 5 $\times$ T1. This design choice helps us balance the MIP running time while providing ample options of target MPs when migrating calls. 

(3) we do not select: (a) calls that have just started (age $<$3 minutes) as participants and media may not have converged (works well as shown in Table \ref{tab:design:duration}), (b) the same call for migration in next iteration of the MIP to reduce repeated interruptions. Such calls are available for migration with a gap of one iteration.

(4) we divide all MPs in a DC into \textit{N virtual clusters}. Each MP is assigned to a cluster randomly. Individual calls on the MPs are unchanged. This way, all calls are also divided across \textit{N} virtual clusters. We run the MIP independently for each of the \textit{N} virtual clusters. As there is no dependencies between clusters, we run MIPs in parallel and roughly speedup the assignment by a factor of \textit{N}. 

\subsection{Migrating calls in multiple waves}
\label{sec:migration:dag}

Once, the MIP calculates the mapping between calls and target MPs, we need to migrate the calls to their target MPs quickly. There are two types of call migration: (a) most of the calls are migrated from hot to cold MPs, which can move simultaneously, (b) in some cases, calls from hot MPs moved to other hot (target) MPs (other calls from such hot target MPs are to be moved before to cool them).

To handle both the cases above, we create a \textit{DAG}. Each stage in the DAG denotes migration step. The nodes in the stage denote the MPs. Each directional edge (e.g., from MP1 to MP2) in the DAG denotes the dependency (e.g., calls on MP1 depend on calls on MP2 to be migrated first). 

We migrate the calls by scrolling backwards in the DAG. The calls are moved to MPs that don't depend on calls on other MPs. When such calls are moved, they free up the capacity on their prior MPs so that the calls from the prior stage can be moved. Our experiments found that, we need at most two levels of DAG.

\section{Evaluation}
\label{sec:eval}

In this section, we evaluate \name through testbed implementation and simulations. The controller implements all five components mentioned in \S\ref{sec:key:arch} using Azure Queues and Redis (more details in \S\ref{sec:eval:bench}) and runs in real time. We replay 24-hour trace collected from one of the \teams DC (detailed in \S\ref{sec:problem}) consisting of millions of calls across 100s of thousands of participants. We replay  the trace where the controller gets signals about participants actions (e.g., joining calls) in the same way as they would in real cases. The \name controller assigns and migrates the calls across MPs in real-time. We calculate the CPU utilization every minute on every MP as detailed in \S\ref{sec:problem:sim}.

Our experiments show: (a) \name substantially reduces hot MPs  as well as calls and participants on the hot MPs (minimum 2.5$\times$, max = 272$\times$), (b) all the three key ideas in \name are effective, (c) benefits of \name continue even at different cluster sizes, (d) increasing calls with apriori knowledge can further reduce hot MPs, (e) predicting max. number of participants has high accuracy, (f) MIP can keep up with the scale due to the optimizations in \S\ref{sec:migration}.

\textbf{Baselines:} We consider 5 baselines:
\begin{enumerate}
    \item Round-robin (RR): This policy simply rotates new calls across MPs as mentioned in \S\ref{sec:problem:sim}.
    \item Random (RN): This policy randomly selects the MP for a new call. This policy is synonymous to hash-based policies used by load balancers built for scale including Ananta\cite{ananta:sigcomm13}, Duet\cite{duet:sigcomm14}, Maglev\cite{maglev:nsdi16}.
    \item Least loaded (LL): LL selects the MP with least load at the time of assignment as mentioned in \S\ref{sec:problem:sim}, 
    \item Least load random (LLR): This policy first selects $K$ least loaded MPs and then selects one MP out of $K$ uniformly at random as mentioned in \S\ref{sec:problem:sim}.
    \item Power-of-two (P2)\cite{power2:itpds01}: assigns calls by selecting 2 MPs at random and then assigning the call to the least loaded MP between the two. P2 is also used by Microsoft's YARP\cite{yarp:web}.
\end{enumerate}

We also use the simple migration algorithm described in \S\ref{sec:problem} in addition to above baseline algorithms. 


\textbf{Setup.} We use a cluster size of 3000 MPs as we want to provide good performance at lower costs. We use COIN-OR\cite{coinor:web} solver for the MIP. We run the MIP every 2 minutes with $L$ = 1000 and $Cap_{m}$ = 75\% in Fig.\ref{fig:algo:lp}, and number of least loaded MPs ($K$) for LLR to 5 to strike a balance between available MPs and their CPU utilization. We set virtual clusters (\S\ref{sec:lp:input}) for MIP to 4.

\S\ref{sec:eval:hot} to \S\ref{sec:eval:ilp} use 1-day trace and \name controller. We compute CPU utilization using module from simulator in \S\ref{sec:problem}. \S\ref{sec:eval:migration} uses the testbed prototype.

\subsection{Reduction in hot MPs, calls, participants}
\label{sec:eval:hot}

\begin{figure}[t]
\centering
\includegraphics[width=0.45\textwidth, page=5]{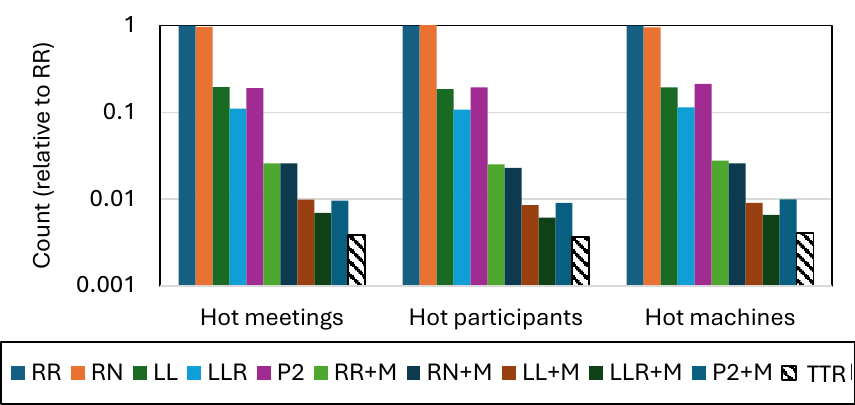}
\caption{Count of hot MPs (machines), and calls and participants on hot MPs (relative to RR). TTR = \name}
\vspace{-0.1in}
\label{fig:eval:hot}
\end{figure}

Fig.\ref{fig:eval:hot} shows the number of hot MPs, calls and participants on such hot MPs. We measure such values every minute and show the aggregate values for 24 hours. The values are normalized to RR. We make the following observations: (a) LLR performs the best among the baselines when not doing migration. It cuts participants on the hot MPs (denoted by $H$) by 9.3$\times$ compared to RR. (b) when using simple migration from \S\ref{sec:problem}, LLR (denoted by LLR+M) again performs the best among the baselines. Compared to RR, it cuts $H$ by 112$\times$. (c) \name performs the best across all algorithms. Compared to RR, it cuts down $H$ by 272$\times$. Compared to LLR+M, \name cuts down $H$ by 2.5$\times$. (d) Random baseline performs similar to RR as both of them select MPs uniformly. 

Above results show improvements in \name in terms of aggregate values over 24 hours. Another important metric is the \textit{max} number of participants on hot MPs at a given time. We found that \name can cut down max. number of participants on hot MPs by 1.42$\times$ compared to LLR+M (not shown). \textit{These results show that \name can significantly improve reliability by reducing the impact of hot nodes.}

\subsection{Reduction in extreme CPU utilization}
\label{sec:eval:cpu}

\begin{figure}[t]
\centering
\includegraphics[width=0.45\textwidth, page=4]{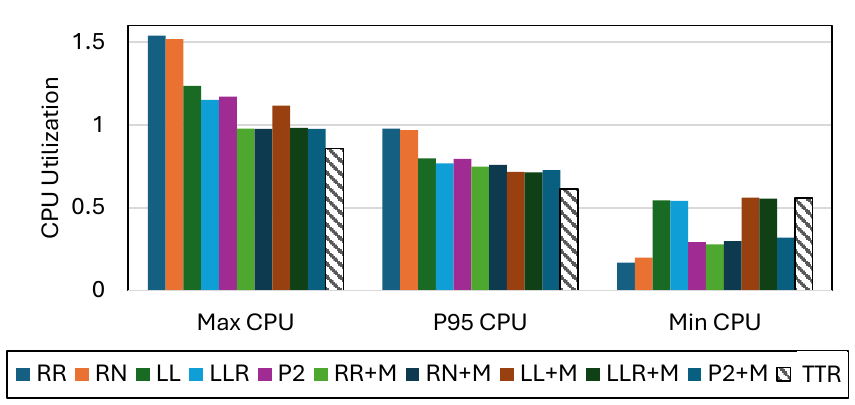}
\caption{CPU utilization (max, P95, and min) across all MPs in the same DC for different policies. +M indicates using simple migration policy from \S\ref{sec:problem}. TTR = \name}
\vspace{-0.1in}
\label{fig:eval:cpu}
\end{figure}


Fig.\ref{fig:eval:cpu} shows the CPU utilization in using different algorithms. We compute max., P95 and min. CPU utilization every minute across all MPs. The max CPU indicates the max. CPU observed across all MPs and all times. Similarly, P95 indicates max. of P95 across all times. Lastly, min. CPU indicates max. of min across all times. Note that the CPU utilization values are above 1 (100\%) for some of the baselines. This is an artifact of the simulations where we add the CPU utilization of all calls on individual MPs.  In reality, such baselines will result in CPU utilization of 100\% but will drop significant number of packets. \textit{We continue to use these definitions in next sections.} We use baseline algorithms without and with migration (denoted by +M). \textit{We find that \name achieves lowest max. CPU and P95 CPU.} Compared to LLR and LLR+M, \name reduces the max. CPU utilization by 34\% and 14\%. 

Recall that MIP in \name tries to keep the CPU utilization bounded to 75\% to keep room for any increase in the CPU due to the new calls placed on the same MPs or new participants joining calls. Consequently, we find that the max. CPU is well under 100\%.


\textbf{CPU utilization imbalance.} Table \ref{tab:problem:cpu} shows the ratio of max. to average CPU utilization for RR, LL and LLR at \textit{busiest} time. We found that P2 performs similarly. For 3000 MPs, such a ratio for P2 is 1.85 and 1.27 without and with migration respectively. In comparison, \name reduces such a ratio to 1.18. \textit{This shows that \name is able to pack the calls more uniformly across MPs.}

\subsection{Impact of different key ideas in \name}
\label{sec:eval:breakdown}

\begin{figure}[t]
\centering
\includegraphics[width=0.45\textwidth, page=3]{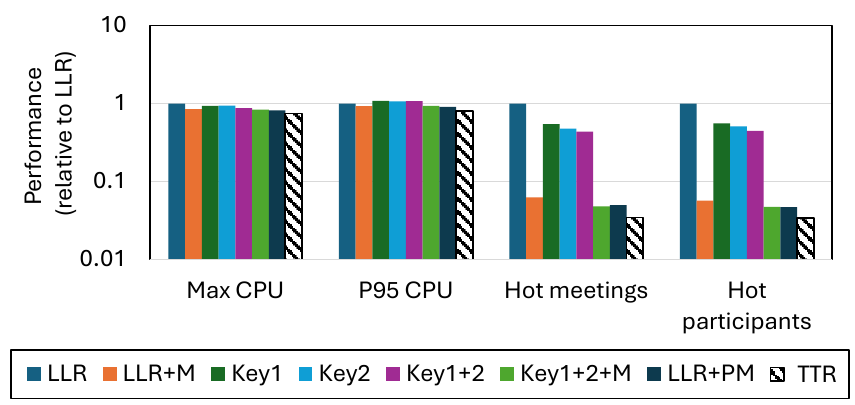}
\caption{Impact of different ideas relative to LLR. TTR = \name.}
\vspace{-0.1in}
\label{fig:eval:breakdown}
\end{figure}

We now turn to evaluate the impact of different key ideas in \name. We evaluate the max. CPU utilization, number of participants on hot MPs in the following cases: (a) baseline of LLR, (b) LLR+M (c) key-1: only changing the initial placement for recurring calls + LLR, (d) key-2: only changing the initial placement for non-recurring calls + LLR (\S\ref{sec:design:ankur}), (e) key1+2, (f) key1+2+M: key1+2 + simple migration strategy from \S\ref{sec:problem}, (g) LLR + PM: baseline LLR + MIP based migration algorithm from \name, (h) \name consisting of all the three key ideas where we use MIP based migration. 

Fig.\ref{fig:eval:breakdown} shows the impact of different cases. All results are normalized to LLR. \textit{It can be seen that key-1 and key-2 substantially improve initial assignments (hot calls and hot participants) compared to LLR.} We see a reduction of 1.8$\times$ and 1.95$\times$ in using key-1 and key-2 over LLR in terms of participants on hot MPs. Combining key-1 and key-2 further improve the performance compared to LLR. Interestingly, the benefits of key-1 and key-2 do not add up when we combine both ideas as they operate on the shared resources (MPs). Next, we see that simple migration strategy clubbed with key1+2 (key1+2+M) can substantially cut down number of participants on hot nodes (21$\times$ over LLR and 1.2$\times$ over LLR+M). \name using all the three key ideas improves by 2$\times$ compared to key1+2+M indicating that MIP based migration algorithm has benefits. Lastly, \name cuts number of participants on hot MPs by 1.43$\times$ over LLR+PM again showing that key ideas 1 and 2 have substantial benefits.


\subsection{Impact of different cluster sizes}
\label{sec:eval:cluster}

\begin{figure}[t]
\centering
\includegraphics[width=0.45\textwidth, page=1]{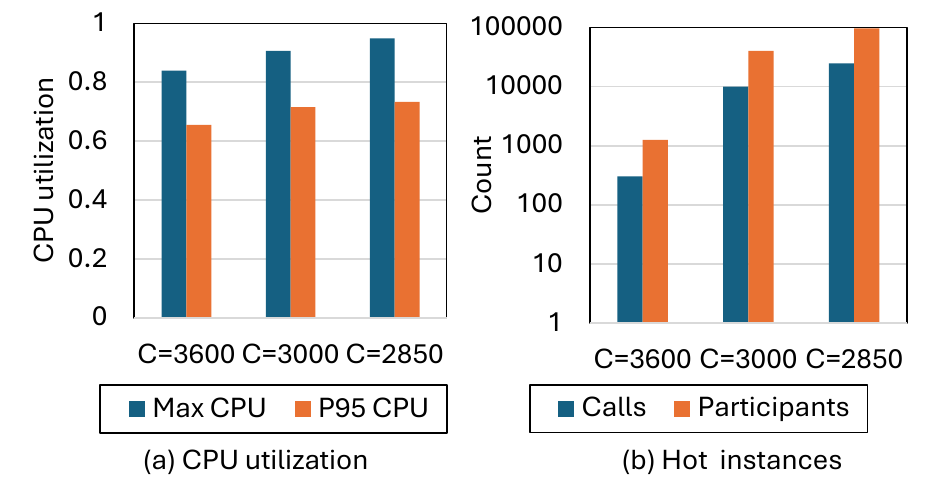}
\caption{Impact of different cluster sizes on \name.}
\vspace{-0.1in}
\label{fig:eval:clustersize}
\end{figure}

In the previous experiments, we kept the cluster size to 3000 MPs. In this section, we evaluate the impact of different cluster sizes. We consider 3 cluster sizes: 3600, 3000 and 2850. Note that the cluster size indicates the total number of MPs and is different from virtual clusters in \S\ref{sec:migration}. Fig.\ref{fig:eval:clustersize} shows the impact of changing cluster sizes on max. and P95 CPU utilization, number of calls and participants on the hot MPs. Fig.\ref{fig:eval:clustersize} shows that as we reduce the cluster size, there is a marginal increase in the max. and P95 CPU utilization. In contrast, reducing cluster size significantly increases number of calls and participants on hot MPs. \textit{Reducing cluster size increases the average CPU utilization and leaves lesser room to pack the calls resulting in higher calls and participants on hot MPs.} Lastly, the average utilization at peak for cluster = 2850 is close to threshold for hot MPs, and we do not suggest reducing cluster size any further.


\subsection{More calls with apriori knowledge}
\label{sec:eval:recurring}

\begin{figure}[t]
\centering
\includegraphics[width=0.45\textwidth, page=1]{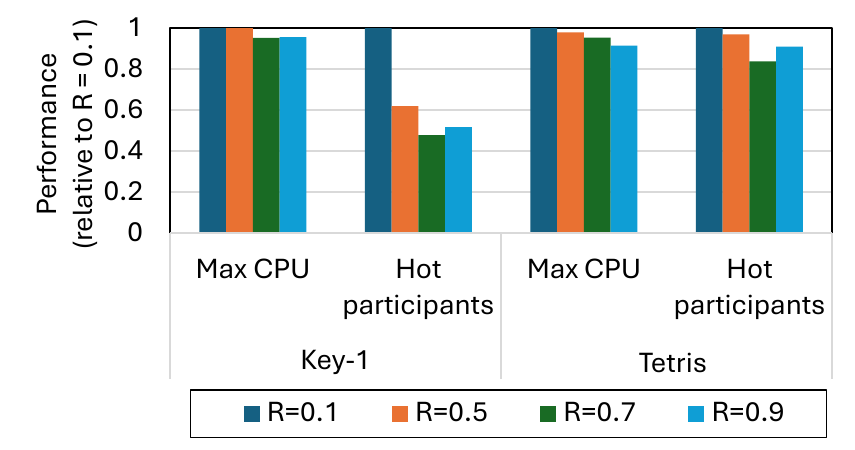}
\caption{Impact of having apriori information about varying fraction of call sizes.}
\vspace{-0.1in}
\label{fig:eval:recurring}
\end{figure}

As mentioned previously, we see 40-60\% calls as part of some recurring call series where we have apriori knowledge about the historical size of the calls. \name does not use any calendar information (e.g., who accepted the call invite). In this experiment, we consider hypothetical cases where we have apriori information about more (and less) fraction of all calls either through recurring series or calendar information. We consider 4 fractions of calls (denoted by $R$) where we have complete knowledge of the calls: $10,\ 50,\ 70,\ 90\%$. We keep cluster size to 3000 MPs, and evaluate impact of $R$ on the first key idea that uses call history as well as \name overall. 


We make three observations: (a) Fig.\ref{fig:eval:recurring} shows that  increasing $R$ significantly helps the key-1  as we can pack better with more knowledge. (b) Increasing $R$ also helps \name. However, the gains are smaller as other two ideas in \name provide benefits even if we do not have knowledge about calls. That said, higher $R$ is still beneficial as number of hot participants reduces with increasing $R$, (c)  Interestingly, $R$ = 90\% performs worse than $R$ = 70\% for key-1 and \name as we are able to pack the calls to reduce the peak CPU utilization. However, it spreads that load on other MPs pushing their CPU utilization above the threshold for the hot MPs.

\subsection{Accuracy of predicting max. participants}
\label{sec:eval:pred}

\begin{figure}[t]
\centering
\includegraphics[width=0.36\textwidth, page=1]{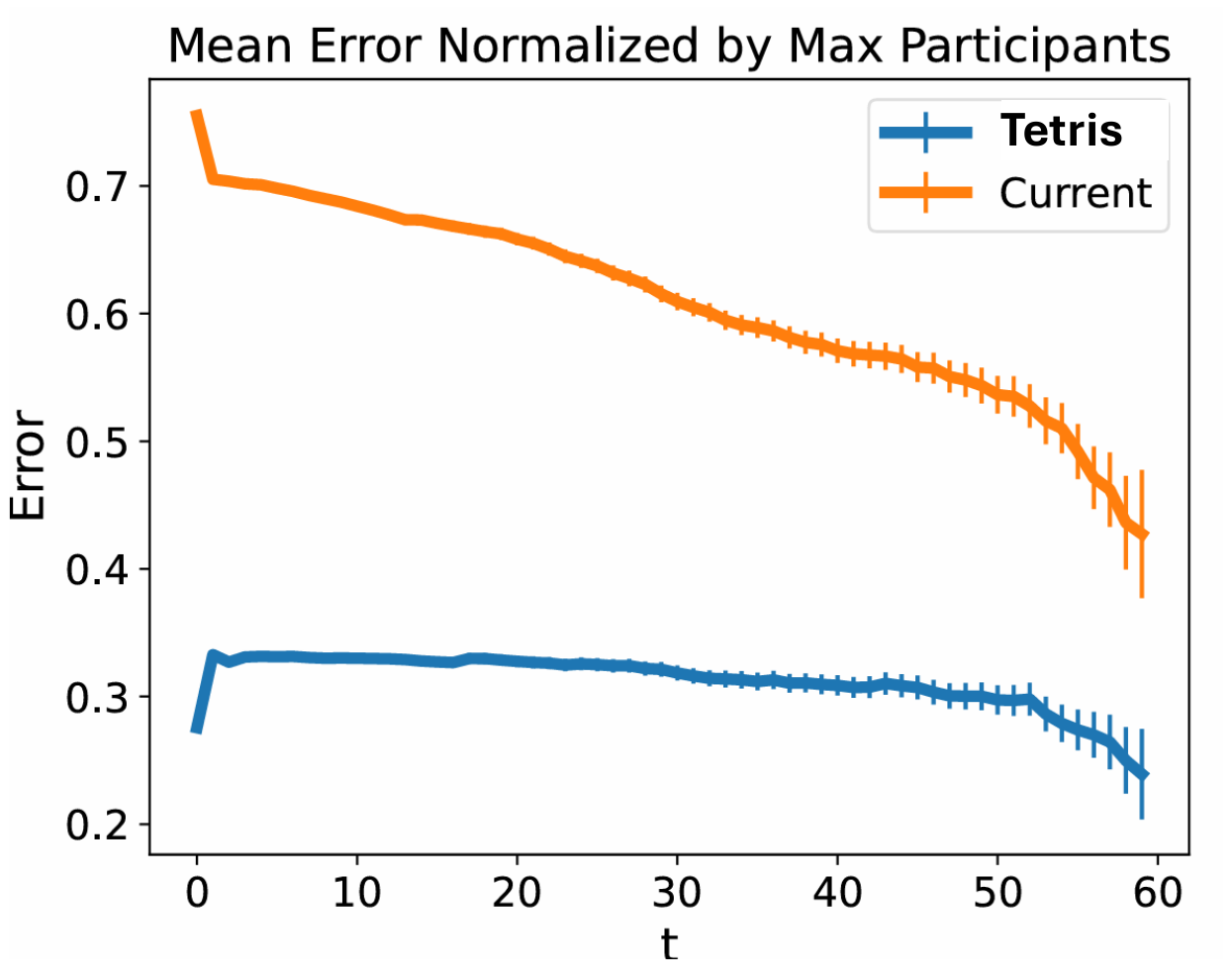}
\caption{Accuracy of predicting peak participants.}
\vspace{-0.15in}
\label{fig:eval:maxpred}
\end{figure}

We show the error in predicting the peak number of participants for non-recurring calls (key idea 2) in Fig.\ref{fig:eval:maxpred}. The X-axis shows time in minutes. The Y-axis shows the normalized error calculated as the absolute difference between predicted and actual peak participants normalized by actual peak participants. The baseline "current" assumes current number of participants as peak. The error bars denote 95$^{th}$ confidence interval. It can be seen that the \textit{prediction algorithm in \name has better accuracy than the baseline and generally improves over time.} In contrast, Fig.\ref{fig:design:stddev} indicates accuracy for recurring calls. 


\subsection{MIP performance}
\label{sec:eval:ilp}

\begin{figure}[t] 
\centering
\includegraphics[width=0.48\textwidth, page=1]{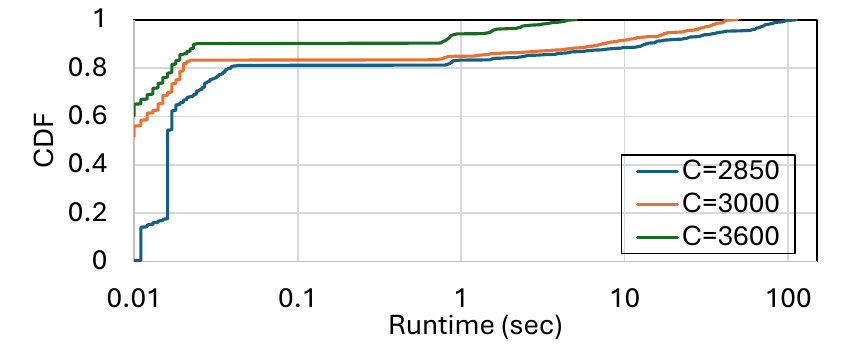}
\caption{MIP Running time for different cluster sizes (C).}
\vspace{-0.1in}
\label{fig:eval:mip}
\end{figure}

We now show the execution time of the MIP. Recall that we run MIP every 2 minutes. When running the MIP, we put the conditions that MIP finishes under 2 minutes or when within 10\% of optimal solution. Fig.\ref{fig:eval:mip} shows the CDF of the running time of the MIP. We show it for \name (using all 3 ideas) when the cluster sizes (C) are 3600, 3000 and 2850. \textit{We found that MIP always produced a migration assignment in all the cases (no failure).} It can be seen that for C = 3600 and 3000, the max. time to compute the migration assignments was 5 and 49 seconds. For cluster size of 2850, max. time bumped up to 107 sec (when the incoming calls are at peak).


\subsection{Testbed: migration performance}
\label{sec:eval:migration}

\begin{SCfigure}[1][t]
  \centering
  \caption{Setup for evaluating impact of call migration.}
  \includegraphics[width=0.30\textwidth, page=1]{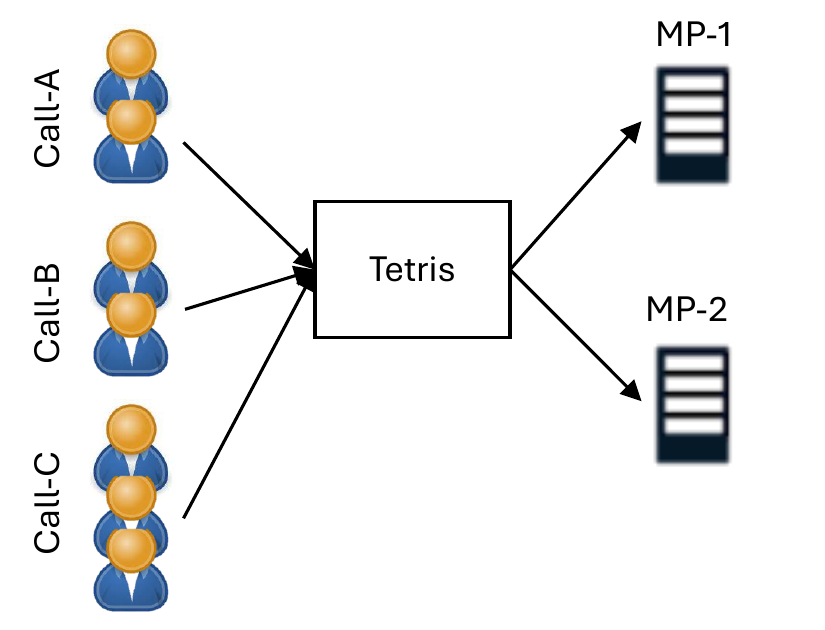}
  \label{fig:eval:testbed:setup}
  \vspace{-0.2in}
\end{SCfigure}

\begin{figure}[t]
\centering
\includegraphics[width=0.4\textwidth, page=1]{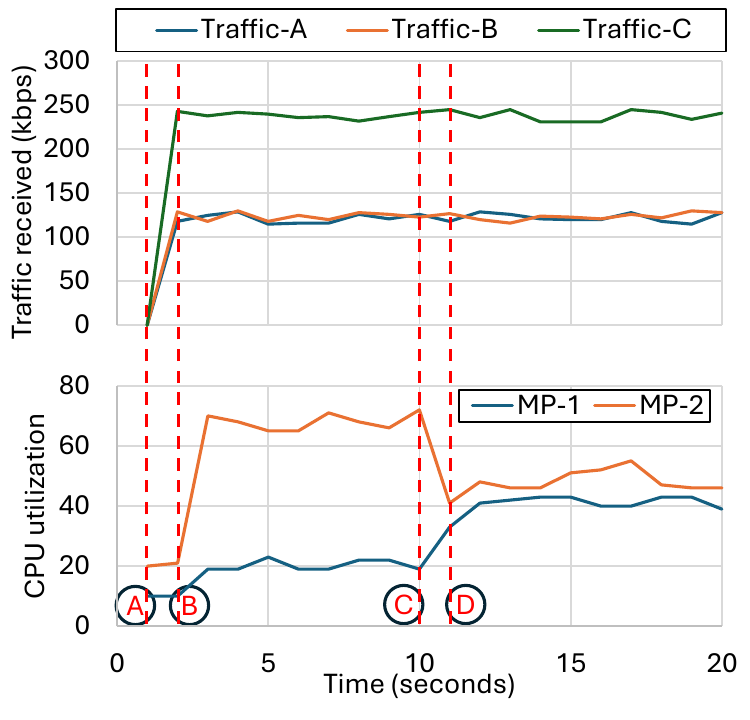}
\caption{CPU util. and traffic received during migration.}
\vspace{-0.1in}
\label{fig:eval:testbed:perf}
\end{figure}

In this section, we show that \name can migrate the calls without any impact on the performance. To do so, we built our MP implementation based on how MPs handle traffic in production (\S\ref{sec:back:mp}). As shown in Fig.\ref{fig:eval:testbed:setup}, we have three calls (A,B and C). Call-A and Call-B have two participants while Call-C has three participants. We have two MPs running on single core VMs (MP-1,2). We first collect traffic trace (UDP packets) for one of the video calls in \teams and replay it at each participant. \name assigns the calls to MPs and later migrates the calls. Fig.\ref{fig:eval:testbed:perf} shows the CPU utilization on two MPs as well as traffic received at one participant of each call. There are four events that we describe below:


\textbf{Event-A.} At time = 1sec, the first participants on all three calls join. Call-A and Call-C are assigned to MP-2 and Call-B is assigned to MP-1. At this point, since only first participants have joined the call, there is no network traffic. \textbf{Event-B.} At time = 2 sec, all the remaining participants join the call; MP-1 and -2 have 2 and 5 participants. The CPU utilization is a function of number of participants (not linear), which is also reflected in CPU utilization (Fig.\ref{fig:eval:testbed:perf}). At this point, the participants start to receive network traffic from other participants on their  calls.


\textbf{Event-C.} to reduce the imbalance in CPU utilization, we migrate call-A to MP-1 starting at time = 10 sec. To do the migration, \name controller sends the signals to MP-1 and MP-2 indicating migration to MP-1. It then sends signals to both the participants of the call-A. However, due to the delays (network or controller), both participants may not start moving to MP-1 at the same time. To account for this, first participant on call-A moves to MP-1 at time = 10 sec, while second participant moves at time = 11 sec (Event-D). During 10$^{th}$ second, MP-2 relayed the traffic to MP-1 for call-A. \textbf{Event-D}. At time = 11 sec, the second participant on call-A moves to MP-1 that completes the call migration. At this point, call-A and call-B are assigned to MP-1, while call-C is assigned to MP-2. MP-1 and MP-2 have 4 and 3 participants and process roughly 1Mbps and 1.1Mbps traffic and CPU utilization changes accordingly (Fig.\ref{fig:eval:testbed:perf}). 

Fig.\ref{fig:eval:testbed:perf} also shows the traffic received by one participant from all three individual calls. We observe that there was no packet drop during migration -- even though both the participants on call-A moved to MP-1 asynchronously. The participants continued to receive video streams from each other and the call progressed without interruptions. Lastly, \teams deploys state-of-art video codecs that are resilient to packet losses to a certain extent and also benefit from jitter buffers\cite{bitag:latency2022}. 
Such mechanisms can help even in the cases of packet loss or reordering during migration. 

\subsection{\name component benchmark}
\label{sec:eval:bench}

\name controller (\S\ref{sec:key:arch}) has many modules. We show the performance and overheads of these modules.

First, \name uses queues to receive messages from individual participants when they join/leave, or when they change the media type or quality. We used Azure Queues\cite{azurequeue:web} to implement such queues, which is equivalent to popular distributed event streaming platform Kafka\cite{kafka:web}. We create the Azure Queue in the same DC as \name controller. We found that we can push messages from different participants at roughly 10K messages/sec. A single thread in \name can only pull roughly 1K messages/sec. However, pulling messages scales with more threads. With 10 threads, \name could keep up with the rate of messages pushed. Currently, a throughput of 10K messages/sec is enough in \name even during busiest time. To increase the throughput further, we simply need to create new queues. Lastly, the cost of using Azure Queues is minuscule compared to number of MPs required in \name.

Second, \name uses Azure Redis\cite{redis:web} to store details about recurring calls. Again, we create Redis in the same DC as \name controller, and observe 1-2 msec read latency. To store number of participants for last 10 instances for 1 billion recurring call series, we only need 38GB (8B call ID + 30B data), and incurs very small cost overhead.

All other components in \name controller run on a single VM with 32 cores. The model that predicts the peak number of participants with the age and current number of participants runs every 24 hours and finishes under a few minutes. It is preloaded in \name controller as key-value pairs and takes $<$1GB. Lastly, we have already covered the latency to run the MIP for migration in \S\ref{sec:eval:ilp}. 

\subsection{Sensitivity analysis}
\label{sec:eval:sense}

\name uses the following parameters. Due to limited space, we explain their sensitivity below:

(1) \textit{Number of instances for LLR ($K$)}: we set it to 5. Lower values converge to LL that suffers from higher hot MPs. Higher values increase the hot MPs slightly as calls get places on MPs with higher CPU. (2) \textit{MIP execution time}: we set it to 2 mins. Lower values can result in MIP not finishing in time. Higher values result in MPs running hot for longer duration. (3) \textit{Number of virtual clusters ($N$)}: We set it to 4. Lower values result in MIP taking longer. Higher values cause resource fragmentation. (4) \textit{Number of migrations ($L$)}: we set it to 1000, which is a small fraction of overall active calls. Larger values result in more migrations but MIP takes shorter time. Smaller values of L reduces number of migrations but make MIP take longer and infeasible at times. E.g., L=1000 results in 16\% higher number of migrations than LLR+M (Fig.\ref{fig:eval:hot}), whereas L=500 results in similar number of migrations as LLR+M but MIP took longer.

\section{Related work}
\label{sec:related}

\name is a novel way to pack the calls within individual DCs. We detail related work here.

\textbf{Conferencing.} Conferencing services such as Zoom\cite{zoom:web}, Microsoft Teams\cite{teams:web}, Google Meet\cite{meet:web}, DingTalk\cite{dingtalk:web}, and others have received considerable community attention\cite{macmillan:imc21, chang:imc21,  carlucci:mmsys16, via:sigcomm16, xron:sigcomm23}. Some of the recent work include: (a) resource management\cite{switchboard:sigcomm23}, (b) network condition based video quality adaptation\cite{gso:sigcomm22}, (c) low latency video transport network\cite{livenet:sigcomm22, lowlatency:sigcomm22}, and (d) codec and transport collaboration\cite{salsify:nsdi18, zhou:mcn19}. In contrast, \name focuses on intelligently packing the calls within DCs to provide good user experience.

\textbf{Layer-4 and layer-7 load balancers (LBs).} Recent works on LB focus on cost, availability, scalability. Ananta\cite{ananta:sigcomm13} and Maglev\cite{maglev:nsdi16} are running in production. \cite{duet:sigcomm14, silkroad:sigcomm17, tiara:nsdi22} use hardware to save costs. \cite{cheetah:nsdi20, beamer:nsdi18} focus on improving the resiliency of LB. Unfortunately, such works focus on spreading TCP flows uniformly that does not work well in \name due to differences in call sizes. Similarly layer-7 LBs such as Yoda\cite{yoda:eurosys16} also fall short as layer-7 information does not contain the call size and hence ends up with poor performance.

\textbf{Server (MP) selection.} Like MP selection problem in \name, prior work to study server/DC/replica selection\cite{taiji:sosp19, fastroute:nsdi15, donar:sigcomm10, liu:ton15, zhang:ieee13, kwon:cloud14}. \cite{c3:nsdi15,shithil:cloud20} focus on replica selection to improve the tail latency. However, such works assign the server only once. In contrast, \name focuses on using call history for initial assignments and migration to cool down hot MPs.

\textbf{VM/container packing.} Many works focus on VM and container packing and migration. \cite{protean:osdi20} and \cite{container:sigkdd22} focus on packing VMs and containers. Unlike \name, such works do not deal with elasticity of the calls -- size of the calls grow and shrink during its lifetime. Also, such works do not have opportunities to use historical data (such as recurring calls). Works such as Borg\cite{borg:eurosys15} focus on preemption. Unfortunately, due to real-time nature of live calls, we cannot preempt the calls.

\textbf{Using apriori knowledge.} Like \name, other works also use apriori knowledge about the workload to improve performance, but operate in different contexts. Corral\cite{virajith:sigcomm15} and Jockey\cite{jockey:eurosys12} show
that 40\% of the big data jobs are recurrent, and use such history to improve performance. Similarly, SLearn\cite{jajoo:nsdi22}, SIA\cite{ishai:socc23}, DOTE\cite{dote:nsdi23} use apriori knowledge to improve CoFlow scheduling, ML scheduling, traffic engineering. \cite{vm:mlsys23, vm:sigmetrics21} predict the lifetime of the VM to improve resource utilization. \name currently does not predict the arrival and lifetime of the calls. We leave it to the future work.

\section{Conclusion}
\label{sec:conc}

We present \name to efficiently pack the calls across MP (Media Processor) servers in \teams -- a large conferencing service. \name splits the assignments in two phases: First, we leverage rich historical call data to improve the call assignment. While it improves the performance, it is not perfect and can still lead to poor user experience for some of the calls. Our second idea is change the MP assignment midway during the calls (once the call size converges) to move the calls away from the hot MPs. Our evaluation using O(10 million) calls from one of our DCs shows \name can reduce number of participants on hot MPs by minimum 2.5$\times$.

\bibliographystyle{abbrv}
\bibliography{references}

\begin{thebibliography}{10}

\bibitem{azurequeue:web}
{Azure Queues}.
\newblock \url{https://learn.microsoft.com/en-us/azure/storage/queues/storage-queues-introduction}.

\bibitem{coinor:web}
{COIN-OR LP solver}.
\newblock \url{https://www.coin-or.org/}.

\bibitem{dingtalk:web}
{Ding Talk}.
\newblock \url{https://www.dingtalk.com}.

\bibitem{meet:web}
{Google Meet}.
\newblock \url{https://apps.google.com/meet}.

\bibitem{kafka:web}
{Kafka distributed event streaming platform}.
\newblock \url{https://kafka.apache.org/}.

\bibitem{lbalgo:web}
{LB DIP selection algorithms}.
\newblock \url{https://www.haproxy.com/solutions/load-balancing/}.

\bibitem{teams:web}
{Microsoft Teams}.
\newblock \url{https://www.microsoft.com/en-us/microsoft-teams/group-chat-software}.

\bibitem{teamsgrowth:web}
{Microsoft Teams user growth}.
\newblock \url{https://www.businessofapps.com/data/microsoft-teams-statistics/}.

\bibitem{yarp:web}
{Microsoft YARP. Yet Another Reverse Proxy.}
\newblock \url{https://microsoft.github.io/reverse-proxy/}.

\bibitem{nginx:web}
{NGINX load balancer}.
\newblock \url{https://docs.nginx.com/nginx/admin-guide/load-balancer/tcp-udp-load-balancer/}.

\bibitem{redis:web}
{Redis in-memory data store}.
\newblock \url{https://redis.io}.

\bibitem{zoom:web}
{Zoom}.
\newblock \url{https://zoom.us/}.

\bibitem{vm:mlsys23}
H.~Barbalho, P.~Kovaleski, B.~Li, L.~Marshall, M.~Molinaro, A.~Pan, E.~Cortez, M.~Leao, H.~Patwari, Z.~Tang, et~al.
\newblock Virtual machine allocation with lifetime predictions.
\newblock {\em Proceedings of MLSys 2023}.

\bibitem{cheetah:nsdi20}
T.~Barbette, C.~Tang, H.~Yao, D.~Kosti{\'c}, G.~Q. Maguire~Jr, P.~Papadimitratos, and M.~Chiesa.
\newblock A high-speed load-balancer design with guaranteed per-connection-consistency.
\newblock In {\em USENIX NSDI 2020}.

\bibitem{bitag:latency2022}
{BITAG}.
\newblock Latency explained.
\newblock Technical report, 2022.
\newblock \url{https://www.bitag.org/documents/BITAG_latency_explained.pdf}.

\bibitem{switchboard:sigcomm23}
R.~Bothra, R.~Gandhi, R.~Bhagwan, V.~N. Padmanabhan, R.~Liang, S.~Carlson, V.~Kamath, S.~Acharyya, K.~Sueda, S.~Chaturmohta, and H.~Sharma.
\newblock Switchboard: Efficient resource management for conferencing services.
\newblock In {\em ACM SIGCOMM 2023}.

\bibitem{vm:sigmetrics21}
N.~Buchbinder, Y.~Fairstein, K.~Mellou, I.~Menache, and J.~S. Naor.
\newblock Online virtual machine allocation with lifetime and load predictions.
\newblock In {\em ACM SIGMETRICS 2021}.

\bibitem{carlucci:mmsys16}
G.~Carlucci, L.~De~Cicco, S.~Holmer, and S.~Mascolo.
\newblock {Analysis and design of the google congestion control for web real-time communication (WebRTC)}.
\newblock In {\em ACM MMSys}, 2016.

\bibitem{chang:imc21}
H.~Chang, M.~Varvello, F.~Hao, and S.~Mukherjee.
\newblock {Can you see me now? A measurement study of Zoom, Webex, and Meet}.
\newblock In {\em ACM IMC}, 2021.

\bibitem{taiji:sosp19}
D.~Chou, T.~Xu, K.~Veeraraghavan, A.~Newell, S.~Margulis, L.~Xiao, P.~M. Ruiz, J.~Meza, K.~Ha, S.~Padmanabha, et~al.
\newblock {Taiji: managing global user traffic for large-scale internet services at the edge}.
\newblock In {\em ACM SOSP}, 2019.

\bibitem{ishai:socc23}
T.~Chugh, S.~Kandula, A.~Krishnamurthy, R.~Mahajan, and I.~Menache.
\newblock Anticipatory resource allocation for ml training.
\newblock In {\em ACM SoCC 2023}.

\bibitem{maglev:nsdi16}
D.~E. Eisenbud, C.~Yi, C.~Contavalli, C.~Smith, R.~Kononov, E.~Mann-Hielscher, A.~Cilingiroglu, B.~Cheyney, W.~Shang, and J.~D. Hosein.
\newblock Maglev: A fast and reliable software network load balancer.
\newblock In {\em USENIX NSDI}, 2016.

\bibitem{jockey:eurosys12}
A.~D. Ferguson, P.~Bodik, S.~Kandula, E.~Boutin, and R.~Fonseca.
\newblock Jockey: guaranteed job latency in data parallel clusters.
\newblock In {\em ACM Eurosys 2012}.

\bibitem{fastroute:nsdi15}
A.~Flavel, P.~Mani, D.~Maltz, N.~Holt, J.~Liu, Y.~Chen, and O.~Surmachev.
\newblock {{FastRoute}: A Scalable {Load-Aware} Anycast Routing Architecture for Modern {CDNs}}.
\newblock In {\em USENIX NSDI}, 2015.

\bibitem{salsify:nsdi18}
S.~Fouladi, J.~Emmons, E.~Orbay, C.~Wu, R.~S. Wahby, and K.~Winstein.
\newblock {Salsify: Low-Latency Network Video through Tighter Integration between a Video Codec and a Transport Protocol}.
\newblock In {\em USENIX NSDI}, 2018.

\bibitem{yoda:eurosys16}
R.~Gandhi, Y.~C. Hu, and M.~Zhang.
\newblock Yoda: a highly available layer-7 load balancer.
\newblock In {\em ACM Eurosys 2016}.

\bibitem{duet:sigcomm14}
R.~Gandhi, H.~Liu, Y.~C. Hu, G.~Lu, J.~Padhye, L.~Yuan, and M.~Zhang.
\newblock {Duet: Cloud Scale Load Balancing with Hardware and Software}.
\newblock In {\em ACM SIGCOMM}, 2014.

\bibitem{knapsacklb:conext25}
R.~Gandhi and S.~Narayana.
\newblock Knapsacklb: Enabling performance-aware layer-4 load balancing.
\newblock volume~3, 2025.

\bibitem{protean:osdi20}
O.~Hadary, L.~Marshall, I.~Menache, A.~Pan, E.~E. Greeff, D.~Dion, S.~Dorminey, S.~Joshi, Y.~Chen, M.~Russinovich, and T.~Moscibroda.
\newblock Protean: Vm allocation service at scale.
\newblock In {\em USENIX OSDI 2020}.

\bibitem{jajoo:nsdi22}
A.~Jajoo, Y.~C. Hu, and X.~Lin.
\newblock A case for task sampling based learning for cluster job scheduling.
\newblock In {\em USENIX NSDI 2022}.

\bibitem{virajith:sigcomm15}
V.~Jalaparti, P.~Bodik, I.~Menache, S.~Rao, K.~Makarychev, and M.~Caesar.
\newblock Network-aware scheduling for data-parallel jobs: Plan when you can.
\newblock In {\em ACM SIGCOMM 2015}.

\bibitem{via:sigcomm16}
J.~Jiang, R.~Das, G.~Ananthanarayanan, P.~A. Chou, V.~Padmanabhan, V.~Sekar, E.~Dominique, M.~Goliszewski, D.~Kukoleca, R.~Vafin, and H.~Zhang.
\newblock {Via: Improving Internet Telephony Call Quality Using Predictive Relay Selection}.
\newblock In {\em ACM SIGCOMM}, 2016.

\bibitem{saving:conext24}
B.~Kataria, P.~LNU, R.~Bothra, R.~Gandhi, D.~Bhattacherjee, V.~N. Padmanabhan, I.~Atov, S.~Ramakrishnan, S.~Chaturmohta, C.~Kotipalli, R.~Liang, K.~Sueda, X.~He, and K.~Hinton.
\newblock Saving private wan: Using internet paths to offload wan traffic in conferencing services.
\newblock In {\em ACM CoNEXT}, 2024.

\bibitem{kwon:cloud14}
M.~Kwon, Z.~Dou, W.~Heinzelman, T.~Soyata, H.~Ba, and J.~Shi.
\newblock {Use of Network Latency Profiling and Redundancy for Cloud Server Selection}.
\newblock In {\em IEEE International Conference on Cloud Computing}, 2014.

\bibitem{livenet:sigcomm22}
J.~Li, Z.~Li, R.~Lu, K.~Xiao, S.~Li, J.~Chen, J.~Yang, C.~Zong, A.~Chen, Q.~Wu, C.~Sun, G.~Tyson, and H.~H. Liu.
\newblock {LiveNet: A Low-Latency Video Transport Network for Large-Scale Live Streaming}.
\newblock In {\em ACM SIGCOMM 2022}.

\bibitem{gso:sigcomm22}
X.~Lin, Y.~Ma, J.~Zhang, Y.~Cui, J.~Li, S.~Bai, Z.~Zhang, D.~Cai, H.~H. Liu, and M.~Zhang.
\newblock {GSO-simulcast: global stream orchestration in simulcast video conferencing systems}.
\newblock In {\em ACM SIGCOMM}, 2022.

\bibitem{liu:ton15}
Z.~Liu, M.~Lin, A.~Wierman, S.~Low, and L.~L.~H. Andrew.
\newblock {Greening Geographical Load Balancing}.
\newblock {\em IEEE/ACM Transactions on Networking}, 2015.

\bibitem{macmillan:imc21}
K.~MacMillan, T.~Mangla, J.~Saxon, and N.~Feamster.
\newblock {Measuring the performance and network utilization of popular video conferencing applications}.
\newblock In {\em ACM IMC}, 2021.

\bibitem{lowlatency:sigcomm22}
Z.~Meng, Y.~Guo, C.~Sun, B.~Wang, J.~Sherry, H.~H. Liu, and M.~Xu.
\newblock Achieving consistent low latency for wireless real-time communications with the shortest control loop.
\newblock In {\em ACM SIGCOMM 2022}.

\bibitem{silkroad:sigcomm17}
R.~Miao, H.~Zeng, C.~Kim, J.~Lee, and M.~Yu.
\newblock {Silkroad: Making stateful layer-4 load balancing fast and cheap using switching asics}.
\newblock In {\em ACM SIGCOMM}, 2017.

\bibitem{power2:itpds01}
M.~Mitzenmacher.
\newblock The power of two choices in randomized load balancing.
\newblock {\em IEEE Transactions on Parallel and Distributed Systems}, 2001.

\bibitem{beamer:nsdi18}
V.~Olteanu, A.~Agache, A.~Voinescu, and C.~Raiciu.
\newblock Stateless datacenter load-balancing with beamer.
\newblock In {\em USENIX NSDI 2018}.

\bibitem{ananta:sigcomm13}
P.~Patel, D.~Bansal, L.~Yuan, A.~Murthy, A.~Greenberg, D.~A. Maltz, R.~Kern, H.~Kumar, M.~Zikos, H.~Wu, et~al.
\newblock {Ananta: Cloud scale load balancing}.
\newblock In {\em ACM SIGCOMM}, 2013.

\bibitem{dote:nsdi23}
Y.~Perry, F.~V. Frujeri, C.~Hoch, S.~Kandula, I.~Menache, M.~Schapira, and A.~Tamar.
\newblock {DOTE}: Rethinking (predictive) {WAN} traffic engineering.
\newblock In {\em USENIX NSDI 2023}.

\bibitem{insidefb:sigcomm15}
A.~Roy, H.~Zeng, J.~Bagga, G.~Porter, and A.~C. Snoeren.
\newblock Inside the social network's (datacenter) network.
\newblock In {\em ACM SIGCOMM 2015}.

\bibitem{shithil:cloud20}
S.~M. Shithil and M.~A. Adnan.
\newblock A prediction based replica selection strategy for reducing tail latency in distributed systems.
\newblock In {\em 2020 IEEE CLOUD}, 2020.

\bibitem{c3:nsdi15}
L.~Suresh, M.~Canini, S.~Schmid, and A.~Feldmann.
\newblock {C3: Cutting tail latency in cloud data stores via adaptive replica selection}.
\newblock In {\em USENIX NSDI}, 2015.

\bibitem{mos:hotnets23}
A.~Taneja, R.~Bothra, D.~Bhattacherjee, R.~Gandhi, V.~N. Padmanabhan, R.~Bhagwan, N.~Natarajan, S.~Guha, and R.~Cutler.
\newblock Don't forget the user: It's time to rethink network measurements.
\newblock In {\em ACM HotNets 2023}.

\bibitem{borg:eurosys15}
A.~Verma, L.~Pedrosa, M.~R. Korupolu, D.~Oppenheimer, E.~Tune, and J.~Wilkes.
\newblock Large-scale cluster management at {Google} with {Borg}.
\newblock In {\em ACM EuroSys 2015}.

\bibitem{donar:sigcomm10}
P.~Wendell, J.~W. Jiang, M.~J. Freedman, and J.~Rexford.
\newblock {DONAR: Decentralized Server Selection for Cloud Services}.
\newblock In {\em ACM SIGCOMM}, 2010.

\bibitem{xron:sigcomm23}
B.~Wu, K.~Qian, B.~Li, Y.~Ma, Q.~Zhang, Z.~Jiang, J.~Zhao, D.~Cai, E.~Zhai, X.~Liu, and X.~Jin.
\newblock Xron: A hybrid elastic cloud overlay network for video conferencing at planetary scale.
\newblock In {\em ACM SIGCOMM 2023}.

\bibitem{container:sigkdd22}
J.~Yan, Y.~Lu, L.~Chen, S.~Qin, Y.~Fang, Q.~Lin, T.~Moscibroda, S.~Rajmohan, and D.~Zhang.
\newblock Solving the batch stochastic bin packing problem in cloud: A chance-constrained optimization approach.
\newblock In {\em ACM SIGKDD 2022}.

\bibitem{tiara:nsdi22}
C.~Zeng, L.~Luo, T.~Zhang, Z.~Wang, L.~Li, W.~Han, N.~Chen, L.~Wan, L.~Liu, Z.~Ding, et~al.
\newblock Tiara: A scalable and efficient hardware acceleration architecture for stateful layer-4 load balancing.
\newblock In {\em USENIX NSDI 2022}.

\bibitem{zhang:ieee13}
Q.~Zhang, Q.~Zhu, M.~F. Zhani, R.~Boutaba, and J.~L. Hellerstein.
\newblock {Dynamic Service Placement in Geographically Distributed Clouds}.
\newblock {\em IEEE Journal on Selected Areas in Communications}, 2013.

\bibitem{zhou:mcn19}
A.~Zhou, H.~Zhang, G.~Su, L.~Wu, R.~Ma, Z.~Meng, X.~Zhang, X.~Xie, H.~Ma, and X.~Chen.
\newblock Learning to coordinate video codec with transport protocol for mobile video telephony.
\newblock In {\em ACM International Conference on Mobile Computing and Networking}, 2019.

\end{thebibliography}

\end{document}